# Data-driven quest for two-dimensional non-van der Waals materials

Rico Friedrich,[1, 2, *] Mahdi Ghorbani-Asl,[1] Stefano Curtarolo,[2, 3] and Arkady V. Krasheninnikov[1, 4]

[1]*Institute of Ion Beam Physics and Materials Research,*
*Helmholtz-Zentrum Dresden-Rossendorf, 01328 Dresden, Germany*
[2]*Center for Autonomous Materials Design, Duke University, Durham, North Carolina 27708, USA*
[3]*Materials Science, Electrical Engineering, and Physics, Duke University, Durham NC, 27708, USA*
[4]*Department of Applied Physics, Aalto University, Aalto 00076, Finland*

Two-dimensional (2D) materials are frequently associated with the sheets that form bulk layered compounds bonded by van der Waals (vdW) forces. The anisotropy and weak interaction between the sheets have also been the main criteria in the computational search for new 2D systems, which predicted about 2000 exfoliable compounds. However, several representatives of a new type of non-vdW 2D systems, such as hematene or ilmenene, which have no layered 3D analogues, and which, unlike, e.g. silicene, do not need to strongly interact with the substrate to be stable, were recently manufactured. The family of non-vdW 2D materials is an attractive playground for data-driven high-throughput approaches as computational design principles are still missing. Here, we outline a new set of 8 binary and 20 ternary candidates by filtering the AFLOW-ICSD database according to the structural prototype of the first two template systems realized in experiment. All materials show a strong structural relaxation upon confinement to 2D which is essential to correctly estimate the bonding strength between facets. The oxidation state of the cations at the surface of the sheets is demonstrated to regulate the inter-facet binding energy with low oxidation states leading to weak bonding. We anticipate this descriptor to be highly useful to obtain novel 2D materials, providing clear guidelines for experiments. Our calculations further indicate that the materials exhibit a vast range of appealing electronic, optical and magnetic properties, which is expected to also make them attractive for potential applications particularly in spintronics.

The isolation of single graphene sheets [1], which proved that two-dimensional (2D) systems can exist, gave rise to the discovery of many 2D materials with unique electronic [2–4], magnetic [5–7], superconducting [8, 9], and topological [10] properties. In addition to being testbeds for studying the behavior of systems in reduced dimensions, 2D materials hold great promise for various applications in opto-electronics [2, 11, 12], catalysis [13, 14] and the energy sector [15–17]. The research effort has been mainly concentrated on the systems which have bulk counterparts representing anisotropic crystals with layers held together by van der Waals (vdW) forces, with the most prominent example being graphene and graphite. The weak interlayer interaction leads to a natural structural separation of the 2D subunits in the crystals, therefore making the mechanical [18] or liquid-phase [19] exfoliation possible.

At the same time, the layered structure allows developing rigorous screening criteria for exfoliability using the entries from large materials databases — outlining data-driven high-throughput investigations as an ideal tool for the discovery and design of 2D compounds [20–28]. Such studies predicted several thousands of vdW 2D systems and were used to set up computational databases [21–23, 25, 26] with the information on their structure and properties.

Unexpectedly, a new direction in the research on 2D systems was opened in 2018, after the experimental realization of 2D materials from non-vdW bonded compounds by a special chemical exfoliation process [29, 30]. The first representatives of these non-vdW 2D materials were hematene and ilmenene obtained from the earth abun-dant ores hematite ($\alpha$-$Fe_2O_3$) and ilmenite ($FeTiO_3$), followed by other systems derived from pyrite ($FeS_2$) [31, 32], chromium sulfide[33], manganese selenide ($\alpha$-$MnSe_2$) [34], metal diborides [35], boro-carbides [36], and diamond-like germanium ($\alpha$-Ge) [37].

While traditional vdW compounds have chemically saturated bonds at the surface of 2D subunits, the aforementioned materials show qualitatively new features. As they are obtained from non-layered systems, they exhibit dangling bonds and surface states, making them more tunable by adsorbates [38]. They have already shown great potential for electronic [39] and optoelectronic [40] applications and by exhibiting enhanced photocatalytic activity for water splitting [29, 30] and photoconductivity [34]. Moreover, they may also offer a versatile playground for magnetism and spintronics in reduced dimensions [41]. First studies mostly focussing on hematene indicate, for instance, tunable magnetic ordering [29, 30, 38, 42–45], as well as spin canting [46].

While there have been significant experimental efforts to generalize recipes for their synthesis [35, 36, 39, 45, 48], computational design principles based on data-driven concepts for outlining non-vdW 2D materials are still missing. It is obvious that the descriptors developed for vdW-bonded systems are not applicable, since they rely on identifying layered subunits in the structure.

Here, we resolve this issue by screening the AFLOW database [49] using the structure of the previously realized representatives as an input and identify 28 non-vdW potentially synthesiseable 2D materials. We further show that they exhibit a vast range of promising electronic, optical, and magnetic properties suggesting in particular spintronic applications. The oxidation state of the cations at the surface of the 2D slab is outlined as a key quan-tity determining the easy exfoliation of the systems thus providing an enabling descriptor for the discovery of novel 2D materials.

* r.friedrich@hzdr.de

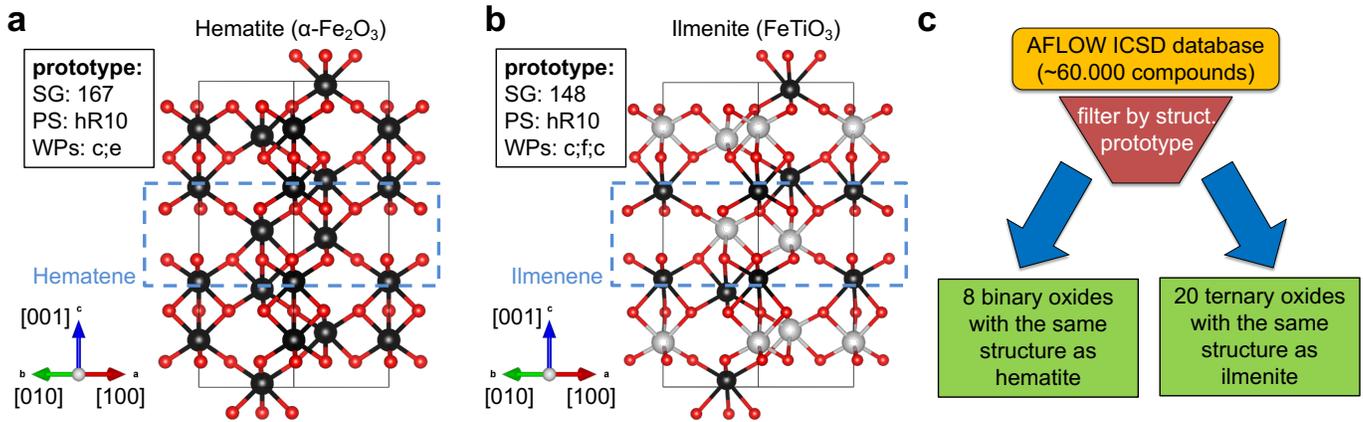

FIG. 1. **Structural prototypes of non-vdW 2D materials.** Atomic structure of (**a**) hematite (α-Fe$_2$O$_3$) and (**b**) ilmenite (FeTiO$_3$), the first non-vdW bulk materials with 2D analogues [29, 30]. For both prototypes, space group (SG), Pearson symbol (PS), and Wyckoff positions (WPs) are indicated in the respective boxes. In each case, the exfoliable [001] facet (monolayer) is indicated in the blue dashed frame leading to hematene and ilmenene 2D materials. Colors: Fe black, O red and Ti light gray [47]. The black line denotes the conventional unit cell. (**c**) Schematic workflow illustrating how the candidate systems were obtained from the AFLOW ICSD database using structural information as an input.

## I. RESULTS AND DISCUSSION

**Outlining more candidates.** Hematite and ilmenite, Fig. 1(**a**) and (**b**), — the first non-vdW materials with 2D analogous [29, 30] — can be used as a template for screening the AFLOW-ICSD database containing ∼60,000 compounds, Fig. 1(**c**). This library primarily incorporates materials realized experimentally. We first assume that the structure of hematite and ilmenite is prone to develop exfoliable units and thus look for other systems with the same structural prototype. Two structures are regarded to have the same prototype if their space group (SG), Pearson symbol (PS) and Wyckoff positions (WPs) match. The structural prototypes [50, 51] of α-Fe$_2$O$_3$ (corundum) and FeTiO$_3$ are indicated in Fig. 1.

The search yields the 8 binary and 20 ternary oxides (excluding P, As and Sb as these are no typical cation species) listed in Tab. I [52]. We decided to focus on oxides since these are abundant materials and typically of high interest for technological applications. For each composition, the listed frequency of occurrence in the database (in brackets after the formula) gives a first indication how common the system is. These candidates are now investigated for the formation of corresponding 2D sheets perpendicular to the [001] direction (called [001] facets) in analogy to the template systems. This facet has been outlined as thermodynamically favorable for hematene [29, 38]. Here, the monolayer structures according to Refs. [29, 30] and Figs. 1(**a**) and (**b**) are considered, which are referred to as 2D systems in the following. This limit provides an upper energetic bound to obtain

2D sheets from a given material [38].

A recent study Ref. [53] demonstrated that the relation between the in-plane tearing energy and the out-of-plane peeling energy controls the length to thickness aspect ratio of liquid-exfoliated nanosheets, implying that layered materials with large mechanical anisotropy prefer to yield nanosheets with large aspect ratios. Similarly, one can expect that non-layered materials with anisotropic bonding schemes can also be exfoliated into non-layered quasi-2D materials according to this assessment [31, 36].

TABLE I. **Binary and ternary compositions with the same structure as hematite and ilmenite.** Binary compositions with the same structural prototype as hematite (α-Fe$_2$O$_3$) and ternary compositions with the same structural prototype as ilmenite (FeTiO$_3$) from the AFLOW-ICSD database. The numbers in brackets after the compound formulas of each system indicate the frequency of occurence in the database. Whether the structure was found to exhibit magnetic moments is also indicated.

| binaries | | ternaries | | | |
|---|---|---|---|---|---|
| composition | mag. | composition | mag. | composition | mag. |
| Al$_2$O$_3$ (72) | no | MgTiO$_3$ (28) | no | CaSnO$_3$ (1) | no |
| Fe$_2$O$_3$ (63) | yes | FeTiO$_3$ (21) | yes | CdGeO$_3$ (1) | no |
| Cr$_2$O$_3$ (52) | yes | GeMgO$_3$ (7) | no | CdTiO$_3$ (1) | no |
| V$_2$O$_3$ (20) | yes | MgSiO$_3$ (7) | no | CoMnO$_3$ (1) | yes |
| Ti$_2$O$_3$ (15) | no | GeMnO$_3$ (6) | yes | CuVO$_3$ (1) | no |
| Rh$_2$O$_3$ (4) | no | MnTiO$_3$ (5) | yes | GeZnO$_3$ (1) | no |
| Ga$_2$O$_3$ (3) | no | NiTiO$_3$ (5) | yes | MnNiO$_3$ (1) | yes |
| In$_2$O$_3$ (3) | no | AgBiO$_3$ (2) | no | SiZnO$_3$ (1) | no |
| | | BiNaO$_3$ (2) | no | SnZnO$_3$ (1) | no |
| | | CoTiO$_3$ (2) | yes | TiZnO$_3$ (1) | no |

**Inter-facet binding energies.** The inter-facet binding energies $\Delta E_b$ computed as the energy difference between relaxed 2D and bulk systems are presented in Fig. 2. We note that the energies are positively defined. Physically, this is the energy required to split a crystal to isolated structural units normalized by the number of units. The term "inter-facet" underscores the point that they are calculated for non-layered materials. They were obtained by relaxing both the ionic positions and cell shape of the 2D sheets. These typically also give a good estimate for the exfoliation energy, *i.e.* the energy needed to peel off one structural 2D unit from the surface of the material [55]. In Figs. S1 and S2, a comparison of the calculated binding energies also including results from different functionals as well as from relaxing different degrees of freedom of the systems is presented showing that these 2D materials primarily gain energy due to the ionic relaxation. For vdW bonded compounds, relaxations of 2D substructures could be omitted when calculating inter-*layer* binding energies [25] which is, however, not a good approximation for the present non-vdW systems. Note that for the ternaries (Fig. 2(**b**)), the 2D slabs can be terminated by either of the two cation species in the formula. The graph includes the results for the energetically more stable termination and the respective terminating element is underlined at the bottom axis. A plot comparing the binding energies obtained with the different terminations is given in Fig. S3.

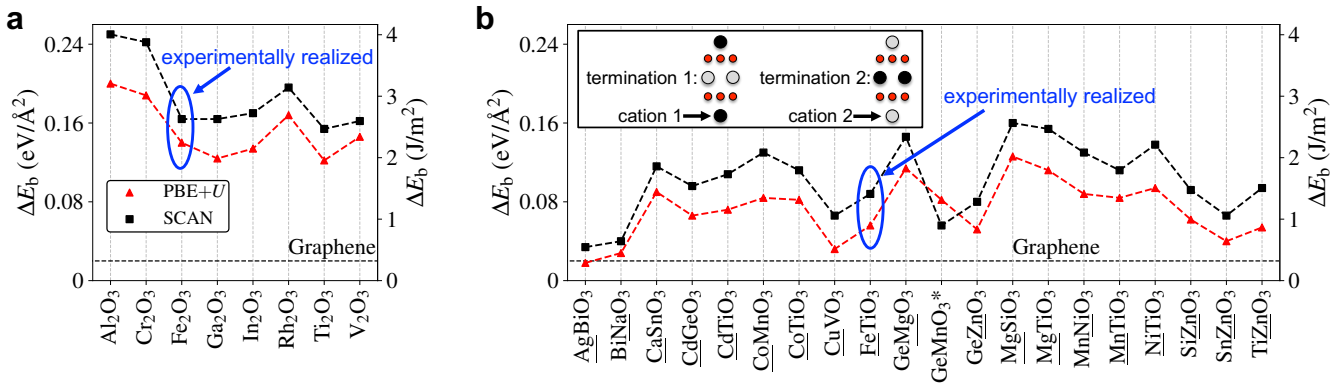

FIG. 2. **Inter-facet binding energies.** Binding energies of investigated (**a**) binary and (**b**) ternary systems for two different functionals. As a reference, the inter-layer binding energy of graphene [25, 54] is indicated by the dashed horizontal black lines. For the ternaries, the data for the slabs with the energetically favourable termination are plotted and the terminating element is underlined at the bottom axis. The inset indicates the two possible terminations. In case of GeMnO₃ (marked by "*"), PBE+$U$ favors Mn termination while for SCAN Ge termination is preferred. Note that in case of Al₂O₃, BiNaO₃, GeMgO₃, and MgSiO₃, PBE+$U$ reduces to plain PBE. Data points for 2D systems realized experimentally [29, 30] are highlighted. The dashed lines connecting the data points are visual guides.

The values are presented for the two DFT methodologies, PBE+$U$ and SCAN. The former is a standard approach for transition metal based systems, which has been used in several previous studies on hematene [29, 38, 42]. The latter is a more sophisticated method, which has been demonstrated to provide accurate structures and energies for diversely bonded systems [56]. Both approximations show similar trends and magnitudes of $\Delta E_b$ with SCAN generally giving higher values. As a reference, the inter-layer binding energy of graphene — well known to be exfoliable — which is found to be $\sim$20 meV/Å² by both theory and experiment [25, 54, 57], is also indicated. While for all binary systems in Fig. 2(**a**) the binding energy is considerably higher than for graphene, the value of 140 meV/Å² computed for hematene with PBE+$U$ nicely fits to previously reported values (after accounting for the different normalization by a factor of two) [38, 42] employing slightly different computational parameters. Considering that hematene has been realized experimentally, all other systems are also likely exfoliable with Al₂O₃ being a potential exception. The systems with the lowest calculated binding energies are close to the limit deemed "potentially exfoliable" (upper bound for $\Delta E_b$ of 130 meV/Å²) in Ref. 25 albeit in this study vdW bonded materials were considered. A limit of $\sim$200 meV/atom for exfoliability has also been proposed [23] which corresponds to about 2/3 of the binding energy calculated for hematene.

As evident from Fig. 2(**b**), the ternaries show in general smaller binding energies in part even close to the graphene reference. Importantly, the value becomes particularly small when the element at the surface of the slab is in a low oxidation state as indicated most prominently by AgBiO₃, BiNaO₃ and CuVO₃ where the surface Ag, Na, and Cu are in state +1. This pattern is related to the fact that smaller surface charges lead to weaker Coulomb like interactions. Also Zn and Fe-termination is energetically favourable, while Mg-termination leads to the highest $\Delta E_b$ among the investigated ternaries, albeit smaller than for hematene, for which exfoliation was achieved experimentally. This general behavior has a strong effect on the magnetic properties, since it results in the magnetic ions Mn²⁺, Fe²⁺, Co²⁺, and Ni²⁺ terminating the 2D materials.

To verify the dynamic stability of the outlined candidates in a high-throughput fashion, 2×2 in-plane supercells are generated from the relaxed structures with the atomic coordinates randomized (gaussian distribution, standard deviation 50 mÅ) [58, 59] and reoptimized. In each case, the systems relax back to the previously found structures.

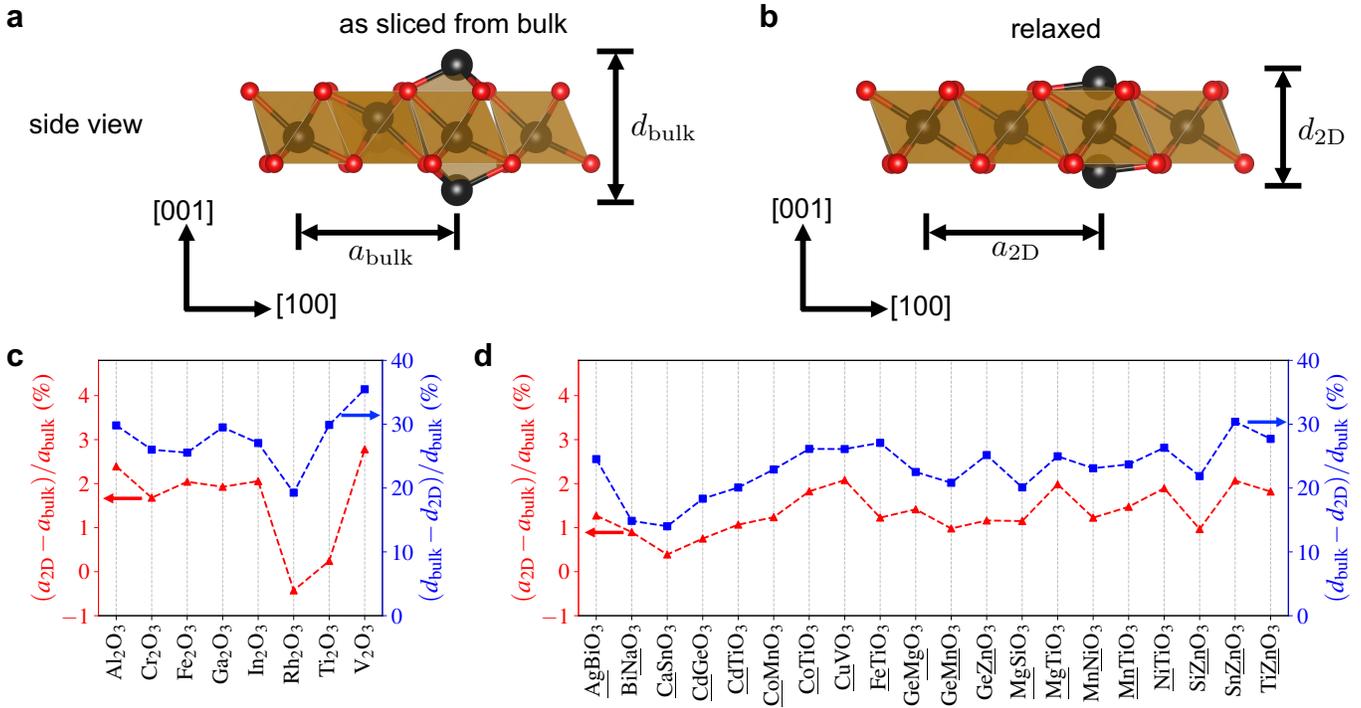

FIG. 3. **Modification of structural parameters.** Side views of the (**a**) initial and (**b**) relaxed unit cell of [001] hematene. Fe coordination polyhedra are visualized in brown. Change of structural parameters for (**c**) binary and (**d**) ternary non-vdW materials upon exfoliation. The modification of the in-plane lattice parameter is indicated by the red curves (left $y$-axis) while the thickness change is given by the blue curves (right $y$-axis). The ternary data are for the energetically preferred slab termination. The dashed lines connecting the data points are visual guides.

**Structural properties.** The investigated non-vdW 2D materials show pronounced structural modifications as compared to their bulk parents. In general, systems compress laterally upon confinement into 2D to compensate for the loss in the out-of-plane bonding by stronger in-plane bonds. The present compounds, however, expand. Figs. 3(**a**) and (**b**) show side views of the initial (as sliced from bulk) and relaxed structures for hematene. The in-plane lattice constants (thickness of the slab) as sliced from the bulk $a_{bulk}$ ($d_{bulk}$) and for the relaxed 2D systems $a_{2D}$ ($d_{2D}$) are also indicated. The relative change of $a_{2D}$ with respect to $a_{bulk}$ is depicted in Figs. 3(**c**) and (**d**) (red curves) for PBE+$U$. An equivalent plot for SCAN showing very similar trends and magnitudes is included in Fig. S4. For all systems, a strong expansion of up to $\sim$3 % for $V_2O_3$ is observed. The only exception is $Rh_2O_3$ for which a small contraction of $\sim -0.5$ % is found.

The expansion correlates well with a corresponding vertical contraction of the systems from $d_{bulk}$ to $d_{2D}$ (blue curves in Figs. 3(**c**) and (**d**)) which ranges up to $\sim$35 % for $V_2O_3$. The mechanism of this behavior can be understood from Figs. 3(**a**) and (**b**). In order to compensate for the dangling bonds created at the surface cations upon exfoliation, the outer atoms move towards the center of the slab during the relaxation reducing the slab thickness. As a consequence, this exerts a stress onto the plain of oxygen anions causing the structures to stretch laterally. For $Rh_2O_3$, the structure compensates for the (smaller) vertical contraction by a twist of the coordination polyhedra rather than a lateral expansion.

**Electronic properties.** The electronic properties of the non-vdW 2D systems are also versatile. A detailed presentation of all band structures and densities of states is shifted to the Supporting Information (Figs. S5 to S32) and here the main focus shall be on the band gaps. Fig. 4 shows the calculated band gap energies $E_{gap}$ for all bulk and 2D systems. While it is well known that DFT — even in the employed PBE+$U$ scheme — in general has problems predicting absolute gap values, the overall trends and differences between similar systems are often still reliable. We thus believe that the observed distribution of the gaps over a wide range (from below 1 eV to almost 6 eV) is noteworthy pointing to the potential usefulness of these systems in *e.g.* optoelectronics. For $Al_2O_3$, $Cr_2O_3$, $Fe_2O_3$, $Rh_2O_3$, and $MgSiO_3$ the 2D gap is considerably smaller than that in the bulk parent system, which can be assigned to the emergence of surface states in the gap upon exfoliation. For the other systems, however, the two gaps are similar or the trend is reversed, *i.e.* the 2D gap is significantly larger than the bulk one as observed for $AgBiO_3$, $FeTiO_3$ and $SnZnO_3$. This is likely associated with the strong structural relaxation described earlier. When the outer cations relax towards the planes of oxygen anions, see Figs. 3(**a**) and (**b**), the electronic interaction is intensified leading to larger level splittings and thus an opening of the gap. This interpretation is corroborated by the fact that for these three systems according to Fig. 3(**d**) a particularly large out-of-plane contraction is found.

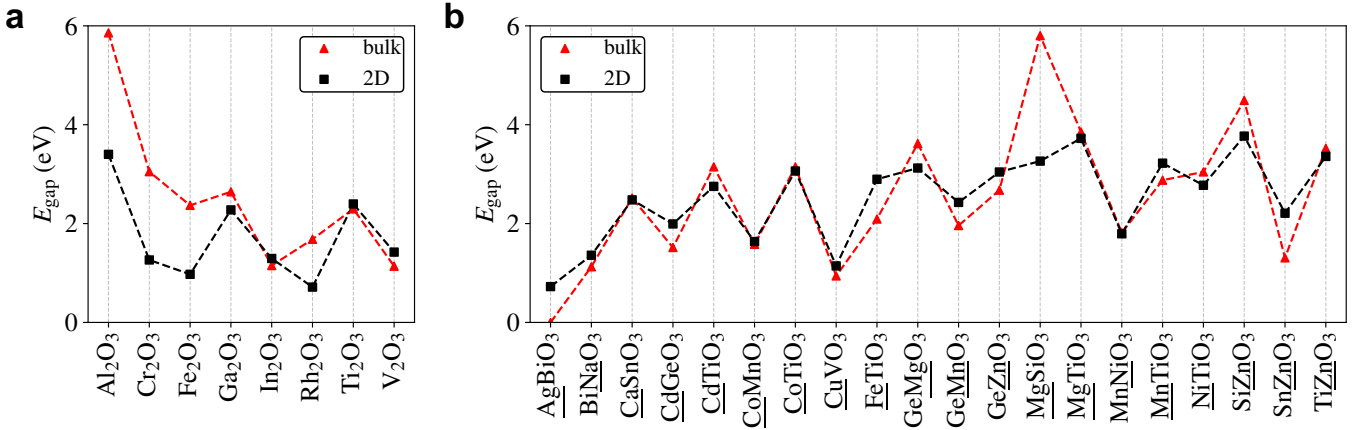

FIG. 4. **Band gap changes upon exfoliation.** Comparison of the calculated band gaps of the bulk and 2D systems for the (**a**) binary and (**b**) ternary compounds.

It shall also be highlighted that interesting linear band crossings, *i.e.* Dirac-like points are observed in the band structures of 2D AgBiO$_3$ and BiNaO$_3$ at the high-symmetry $K$-point warranting further investigations on the potential topological nature of these systems and other non-vdW 2D materials.

**Magnetism.** These mainly transition-metal-based 2D systems are especially interesting in terms of their magnetic properties, since in isotropic crystals magnetic order is prohibited in 2D at finite temperatures according to the Mermin-Wagner-Berezinskii theorem [60]. Finite anisotropy can, however, lift this constraint leading to potential magnetic order, as already reported for hematene and ilmenene [29, 30]. Table I also indicates for which of the bulk parent systems a finite magnetic moment is reported in the AFLOW database. Hence 4 (7) of 8 (20) binary (ternary) systems are found to be magnetic. While the standard workflow [61] of AFLOW might be biased towards ferromagnetic (FM) configurations with small magnetic moments, we have further investigated the preferred magnetic ordering according to the algorithm developed within the coordination corrected enthalpies (CCE) method [62, 63] (see Methods section for further information). This ansatz reliably finds the antiferromagnetic (AFM) configurations of hematite and hematene reported previously [38]. Note that for the 2D systems, the size of the moments of the magnetic ions terminating the slab (outer ions) is generally not equal to the ones within the slab (inner ions) and hence the AFM configuration might have a small net moment as they are not fully compensated. This is particularly evident for ternary systems with two different magnetic ions such as CoMnO$_3$, and MnNiO$_3$.

The spin configurations found are indicated in Fig. 5 and additional results including the energy difference $\Delta E = E_{AFM} - E_{FM}$ between the lowest energy AFM and FM ordering and magnetic moments also including bulk reference data are summarized in Table II. Results for SCAN are included in section VI. in the Supporting Information. The 2D materials are primarily AFM with the exception of Cr$_2$O$_3$ which becomes FM (Fig. 5(**a**)) upon confinement to 2D as noted previously albeit for oxygen terminated slabs [44]. For Fe$_2$O$_3$, the moments of the inner ions are anti-aligned to the ones at bottom and top, Fig. 5(**b**). On the contrary, for Ti$_2$O$_3$ and V$_2$O$_3$ also the inner moments show AFM ordering resulting in the top and bottom moments being anti-aligned (Fig. 5(**c**)). For ternaries, only two types of magnetic configurations are observed (Figs. 5(**d**) and (**e**)). If two magnetic species are present as for CoMnO$_3$ and MnNiO$_3$, the inner moments are AFM to the outer ones. For the other cases with only one magnetic species, the spins at top and bottom are AFM. The moments of the ions at the surface of the slab generally reduce by up to $\sim 0.2 \ \mu_B$ when compared to their bulk reference values. The only exception is Cr$_2$O$_3$, for which they are slightly enhanced.

Even with most systems being AFM, the general behavior that the species in lower oxidation state is at the surface of the slabs, as mentioned before, is very appealing. It causes the magnetic ions of the ternary systems *i.e.* Mn$^{2+}$, Fe$^{2+}$, Co$^{2+}$, and Ni$^{2+}$ to be at the surface of the 2D materials. This outlines an ideal playground for spintronics since the spin polarization at the surfaces can be accessed by techniques such as spin polarized scanning tunneling microscopy (STM) and tuned by *e.g.* adsorbates [38, 64]. While the 2D systems themselves are found to be insulating, it has already been demonstrated that, due to weak interactions, the electronic and magnetic properties of hematene were well preserved on Au(111) [43] — a prototypical substrate in surface science.

To underscore this point, Fig. 6 presents the surface spin polarization on an isosurface of the charge density for CoTiO$_3$, GeMnO$_3$, and MnTiO$_3$. We decided to use the total charge density to clearly indicate that the whole density is strongly spin polarized and not just a part of it within a certain energy interval (as sampled by STM). For this, we have studied the FM and AFM alignment of the surface spins in a $2 \times 2$ supercell as the inter-plane couplings (see Fig. 5) are regarded to be reliably determined from the study of the structural unit cell. For GeMnO$_3$, the surface moments couple antiferromagnetically whereas for the other systems FM order is observed. As indicated in Fig. 6, for CoTiO$_3$, a positive polarization reaching up to $\sim 70$ % in a narrow region above the Co centers is observed. GeMnO$_3$ depicts a strong spatial variation of the polarization from $-86$ % to $+86$ % due to the AFM in-plane coupling. Finally, MnTiO$_3$ shows high positive values extending over a larger spatial range above the Mn ions compared to CoTiO$_3$ also reaching a maximum of $+86$ %. These examples showcase the versatile degree of surface spin polarization that can be expected from non-vdW 2D materials offering great potential for spintronic applications.

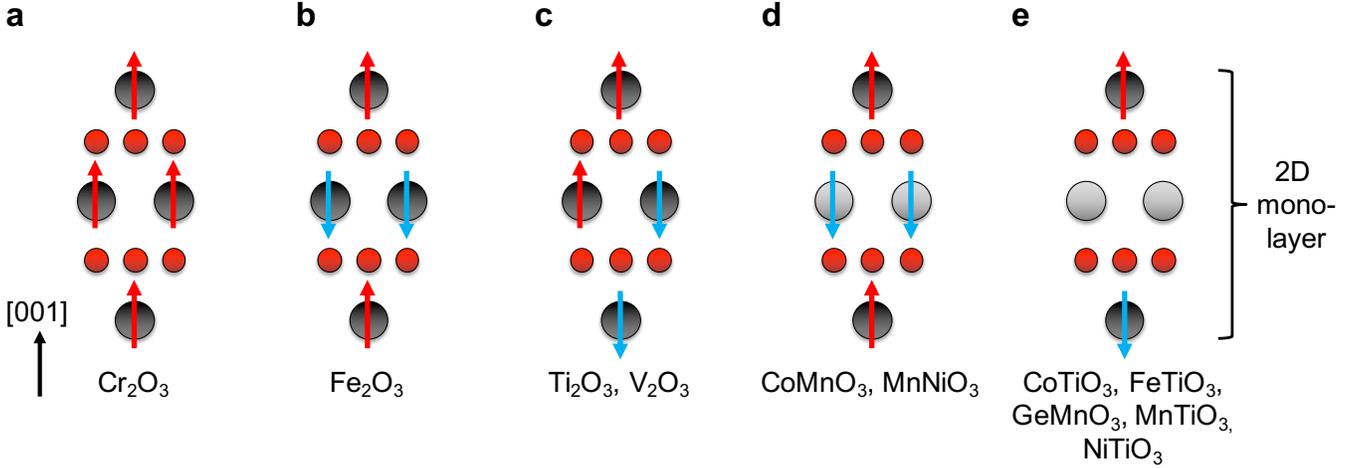

FIG. 5. **Magnetic configurations.** Schematic representation of the magnetic configurations for (**a-c**) binary and (**d** and **e**) ternary 2D systems. The compositions for which the specific configuration applies are indicated at the bottom. Cations are indicated by black and light gray spheres, whereas oxygen anions are in red.

TABLE II. **Magnetic properties.** Energetically preferred magnetic ordering type, energy difference $\Delta E = E_{\mathrm{AFM}} - E_{\mathrm{FM}}$ between lowest energy AFM and FM ordering in eV/formula unit, and absolute magnetic moments for the inner and outer magnetic ions of the slab in $\mu_{\mathrm{B}}$ for the bulk and corresponding [001] facets. For ternaries, the $A$ element is the first cation species in the formula while the $B$ element corresponds to the second. The terminating elements of the slabs (outer ions) are underlined.

| formula | mag. order type | | $\Delta E$ | | $\mu_{\mathrm{bulk}}$ | $\mu_{\mathrm{2D}}$ | |
|---|---|---|---|---|---|---|---|
| | bulk | 2D | bulk | 2D | | inner | outer |
| Cr$_2$O$_3$ | AFM | FM | −0.044 | 0.013 | 2.91 | 2.97 | 3.04 |
| Fe$_2$O$_3$ | AFM | AFM | −0.489 | −0.479 | 4.18 | 4.20 | 4.01 |
| Ti$_2$O$_3$ | AFM | AFM | −0.893 | −0.094 | 0.87 | 0.89 | 0.74 |
| V$_2$O$_3$ | AFM | AFM | −0.013 | −0.090 | 1.83 | 1.84 | 1.67 |

| formula | mag. order type | | $\Delta E$ | | $\mu_{\mathrm{bulk}}$ | | $\mu_{\mathrm{2D}}$ | |
|---|---|---|---|---|---|---|---|---|
| | bulk | 2D | bulk | 2D | $A$ el. | $B$ el. | inner | outer |
| C̲o̲MnO$_3$ | AFM | AFM | −0.024 | −0.062 | 2.77 | 3.07 | 3.12 | 2.72 |
| C̲o̲TiO$_3$ | FM | AFM | 0.004 | −0.011 | 2.79 | 0.06 | 0.00 | 2.74 |
| F̲e̲TiO$_3$ | FM | AFM | 0.003 | −0.007 | 3.72 | 0.00 | 0.00 | 3.69 |
| G̲e̲MnO$_3$ | AFM | AFM | −0.041 | −0.014 | 0.01 | 4.61 | 0.00 | 4.57 |
| Mn̲N̲i̲O$_3$ | AFM | AFM | −0.054 | −0.147 | 3.07 | 1.73 | 3.13 | 1.69 |
| M̲n̲TiO$_3$ | AFM | AFM | −0.027 | −0.018 | 4.60 | 0.02 | 0.00 | 4.55 |
| N̲i̲TiO$_3$ | FM | AFM | 0.001 | −0.001 | 1.75 | 0.07 | 0.00 | 1.73 |

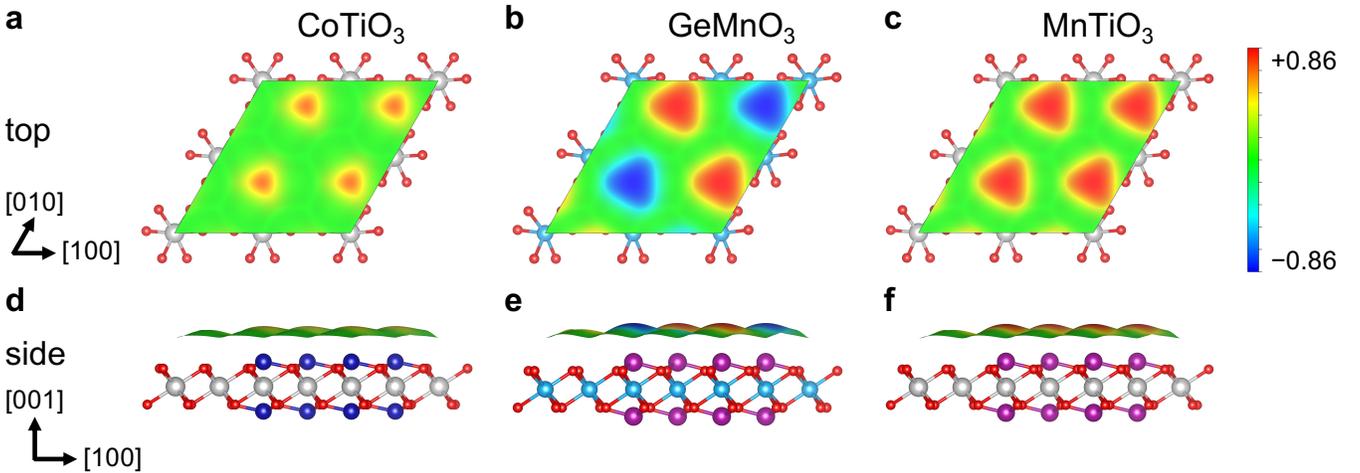

FIG. 6. **Surface spin polarization.** Surface spin polarization on an isosurface of the charge density for CoTiO$_3$ (**a** and **d**), GeMnO$_3$ (**b** and **e**), and MnTiO$_3$ (**c** and **f**) in top and side view, respectively. The isovalue for the charge density is $3 * 10^{-4}$ $e$/Å$^3$. The scale bar indicates the degree of spin polarization. Colors: Co dark blue, Ti light gray, Ge light blue, Mn magenta, and O red.

## II. CONCLUSIONS

We have outlined a new set of non-vdW 2D materials by employing data-driven concepts and high-throughput calculations. By filtering the AFLOW-ICSD database according to the structural prototype of the two first template systems realized experimentally, we have obtained 8 binary and 20 ternary candidates. Although the calculated inter-facet binding energies of several systems are considerably higher than for the graphene reference, they are all lower or comparable to the ones of the first manufactured non-vdW 2D systems presenting thus no hurdle for their realization. The oxidation number of the cations at the surface of the 2D slab is identified as a suitable descriptor for exfoliability, $i.e.$, indicating easy exfoliation in case of a low oxidation state — a principle of likely high value for future 2D materials discovery. In terms of the structure, the 2D systems show a strong vertical contraction and lateral expansion as compared to the bulk parents. The band gaps are distributed over a large range and potential topological features are exhibited by several candidates. The magnetic properties are especially appealing: while most systems show AFM ordering, the magnetic ions are at the surface, which leads to a very diverse set of surface spin polarisations, foreshadowing potential applications in spintronics. We anticipate that our study will prove useful for the discovery of new non-vdW 2D systems and will unravel the potential of this class of novel materials.

## METHODS

The $ab$-$initio$ calculations for the exchange-correlation functionals LDA [65, 66], PBE [67], SCAN [68], and PBE+$U$ [69–71] are performed with AFLOW [72, 73] and the Vienna $Ab$-$initio$ Simulation Package (VASP) [74–76] with settings according to the AFLOW standard [61] and the internal VASP precision set to ACCURATE. For calculations with SCAN, projector-augmented-wave (PAW) pseudopotentials [77] of VASP version 5.4 are used and non-spherical contributions to the gradient of the density in the PAW spheres are explicitly included for SCAN and PBE+$U$. The [001] monolayer 2D facets are constructed from the bulk standard conventional unit cell with the respective AFLOW commands [78] resulting in structures with 10 atoms and at least 20 Å of vacuum perpendicular to the slabs are included. For the facets, both the ionic positions and the cell shape are allowed to relax unless stated otherwise. The AFLOW internal automatic determination of $k$-point sets is used and for the calculations of the 2D facets, the setting for the number of $k$-points per reciprocal atom [61] is reduced to 1,000 resulting in $\Gamma$-centered $10 \times 10 \times 1$ grids.

The bulk and 2D candidate systems containing potentially magnetic elements such as Ti, V, Cr, Mn, Fe, Co, Ni, and Rh are rigorously checked for magnetism using the algorithm developed within the coordination corrected enthalpies (CCE) method [62], $i.e.$ investigating all possible FM and AFM configurations in the structural unit cell for five different sizes of induced magnetic moments each. The analysis is only applied to other systems when the standard workflow of AFLOW [61] resulted in finite magnetic moments after the relaxation. Ferrimagnetic configurations, $i.e.$ having one moment anti-aligned to the other three in the unit cell, are only checked for the bulk binary systems but never resulted in the lowest energy configuration and are hence not considered for the 2D facets and ternary systems. In each case, the lowest energy magnetic state is used for the further calculations.

The spin polarization is defined as:

$$P = \frac{n_\uparrow - n_\downarrow}{n_\uparrow + n_\downarrow}, \qquad (1)$$

where $n_\uparrow$ and $n_\downarrow$ correspond to up spin and down spin densities, respectively. For the relaxed 2D systems, a static electronic calculation on $2 \times 2$ supercells was carried out. FM and AFM in-plane magnetic ordering was considered with the size of the magnetic moments induced according to the lowest energy magnetic configuration found for the structural unit cell. In each case, due to the weak in-plane magnetic coupling, the forces on the atoms and stress on the supercell were found to be negligibly small in each case.

The inter-facet binding energy is computed as:

$$\Delta E_b = \frac{E_{slab} - E_{bulk}}{A}, \qquad (2)$$

where $E_{slab}$ and $E_{bulk}$ indicate the total energies of the relaxed 2D material and bulk, respectively and $A$ is the surface area according to the relaxed bulk unit cell. As pointed out in Ref. 55 for vdW 2D systems, it also gives a good estimate of the exfoliation energy from the surface. We believe that this correspondence is still valid in case of non-vdW 2D systems since — although atomic relaxations are much larger — these are likely comparable when increasing the inter-facet distance to estimate the binding energy and when peeling a single facet from the surface.

Numerical data for the inter-facet binding energies, structural parameters and magnetic properties are included in section VI. in the Supporting Information.

## ASSOCIATED CONTENT

### Supporting Information

Inter-facet binding energies for additional functionals and from relaxing different degrees of freedom, inter-facet binding energies for different terminations for ternaries, change of structural parameters for SCAN, band structures for all binaries and ternaries, and tables with numerical data.

## ACKNOWLEDGMENTS

The authors thank the HZDR Computing Center, HLRS, Stuttgart, Germany and TU Dresden Cluster "Taurus" for generous grants of CPU time. R.F. acknowledges support from the Alexander von Humboldt foundation under the Feodor Lynen research fellowship. A.V.K. thanks the German Research Foundation (DFG), project KR 4866/2-1 for the support. R.F. thanks Marco Esters, Corey Oses, David Hicks, Silvan Kretschmer, and Xiomara Campilongo for fruitful discussions.

[1] K. S. Novoselov, A. K. Geim, S. V. Morozov, D. Jiang, Y. Zhang, S. V. Dubonos, I. V. Grigorieva, and A. A. Firsov, *Electric Field Effect in Atomically Thin Carbon Films*, Science **306**, 666–669 (2004).

[2] S. Z. Butler, S. M. Hollen, L. Cao, Y. Cui, J. A. Gupta, H. R. Gutiérrez, T. F. Heinz, S. S. Hong, J. Huang, A. F. Ismach, E. Johnston-Halperin, M. Kuno, V. V. Plashnitsa, R. D. Robinson, R. S. Ruoff, S. Salahuddin, J. Shan, L. Shi, M. G. Spencer, M. Terrones, W. Windl, and J. E. Goldberger, *Progress, Challenges, and Opportunities in Two-Dimensional Materials Beyond Graphene*, ACS Nano **7**, 2898–2926 (2013).

[3] M. Chhowalla, H. S. Shin, G. Eda, L.-J. Li, K. P. Loh, and H. Zhang, *The chemistry of two-dimensional layered transition metal dichalcogenide nanosheets*, Nature Chemistry **5**, 263–275 (2013).

[4] S. Manzeli, D. Ovchinnikov, D. Pasquier, O. V. Yazyev, and A. Kis, *2D transition metal dichalcogenides*, Nature Reviews Materials **2**, 1–15 (2017).

[5] K. S. Burch, D. Mandrus, and J.-G. Park, *Magnetism in two-dimensional van der Waals materials*, Nature **563**, 47–52 (2018).

[6] M. Gibertini, M. Koperski, A. F. Morpurgo, and K. S. Novoselov, *Magnetic 2D materials and heterostructures*, Nat. Nanotechnol. **14**, 408–419 (2019).

[7] Y. L. Huang, W. Chen, and A. T. S. Wee, *Two-dimensional magnetic transition metal chalcogenides*, SmartMat **2**, 139–153 (2021).

[8] Y. Cao, V. Fatemi, S. Fang, K. Watanabe, T. Taniguchi, E. Kaxiras, and P. Jarillo-Herrero, *Unconventional superconductivity in magic-angle graphene superlattices*, Nature **556**, 43–50 (2018).

[9] D. Campi, S. Kumari, and N. Marzari, *Prediction of Phonon-Mediated Superconductivity with High Critical Temperature in the Two-Dimensional Topological Semimetal W2N3*, Nano Lett. **21**, 3435–3442 (2021).

[10] L. Kou, Y. Ma, Z. Sun, T. Heine, and C. Chen, *Two-Dimensional Topological Insulators: Progress and Prospects*, J. Phys. Chem. Lett. **8**, 1905–1919 (2017).

[11] Q. H. Wang, K. Kalantar-Zadeh, A. Kis, J. N. Coleman, and M. S. Strano, *Electronics and optoelectronics of two-dimensional transition metal dichalcogenides*, Nat. Nanotechnol. **7**, 699–712 (2012).

[12] M. C. Lemme, L.-J. Li, T. Palacios, and F. Schwierz, *Two-dimensional materials for electronic applications*, MRS Bull. **39**, 711–718 (2014).

[13] D. Deng, K. S. Novoselov, Q. Fu, N. Zheng, Z. Tian, and X. Bao, *Catalysis with two-dimensional materials and their heterostructures*, Nat. Nanotechnol. **11**, 218–230 (2016).

[14] Y. Wang, J. Mao, X. Meng, L. Yu, D. Deng, and X. Bao, *Catalysis with Two-Dimensional Materials Confining Single Atoms: Concept, Design, and Applications*, Chem. Rev. **119**, 1806–1854 (2019).

[15] B. Anasori, M. R. Lukatskaya, and Y. Gogotsi, *2D metal carbides and nitrides (MXenes) for energy storage*, Nature Reviews Materials **2**, 1–17 (2017).

[16] J. Wang, V. Malgras, Y. Sugahara, and Y. Yamauchi, *Electrochemical energy storage performance of 2D nanoarchitectured hybrid materials*, Nat. Commun. **12**, 3563 (2021).

[17] I. V. Chepkasov, M. Ghorbani-Asl, Z. I. Popov, J. H. Smet, and A. V. Krasheninnikov, *Alkali metals inside bilayer graphene and MoS2: Insights from first-principles calculations*, Nano Energy **75**, 104927 (2020).

[18] K. S. Novoselov, D. Jiang, F. Schedin, T. J. Booth, V. V. Khotkevich, S. V. Morozov, and A. K. Geim, *Two-dimensional atomic crystals*, Proc. Natl. Acad. Sci. **102**, 10451–10453 (2005).

[19] V. Nicolosi, M. Chhowalla, M. G. Kanatzidis, M. S. Strano, and J. N. Coleman, *Liquid Exfoliation of Layered Materials*, Science **340**, 1226419 (2013).

[20] E. S. Penev, N. Marzari, and B. I. Yakobson, *Theoretical Prediction of Two-Dimensional Materials, Behavior, and Properties*, ACS Nano **15**, 5959–5976 (2021).

[21] S. Lebègue, T. Björkman, M. Klintenberg, R. M. Nieminen, and O. Eriksson, *Two-Dimensional Materials from Data Filtering and Ab Initio Calculations*, Phys. Rev. X **3**, 031002 (2013).

[22] F. A. Rasmussen and K. S. Thygesen, *Computational 2D Materials Database: Electronic Structure of Transition-Metal Dichalcogenides and Oxides*, J. Phys. Chem. C **119**, 13169–13183 (2015).

[23] K. Choudhary, I. Kalish, R. Beams, and F. Tavazza, *High-throughput Identification and Characterization of Two-dimensional Materials using Density functional theory*, Sci. Rep. **7**, 5179 (2017).

[24] G. Cheon, K.-A. N. Duerloo, A. D. Sendek, C. Porter, Y. Chen, and E. J. Reed, *Data Mining for New Two-and One-Dimensional Weakly Bonded Solids and Lattice-Commensurate Heterostructures*, Nano Lett. **17**, 1915–1923 (2017).

[25] N. Mounet, M. Gibertini, P. Schwaller, D. Campi, A. Merkys, A. Marrazzo, T. Sohier, I. E. Castelli, A. Cepellotti, G. Pizzi, and N. Marzari, *Two-dimensional materials from high-throughput computational exfoliation of experimentally known compounds*, Nat. Nanotechnol. **13**, 246–252 (2018).

[26] R. Besse, M. P. Lima, and J. L. F. Da Silva, *First-Principles Exploration of Two-Dimensional Transition Metal Dichalcogenides Based on Fe, Co, Ni, and Cu Groups and Their van der Waals Heterostructures*, ACS Appl. Energy Mater. **2**, 8491–8501 (2019).

[27] M. Núñez, R. Weht, and M. Núñez-Regueiro, *Searching for electronically two dimensional metals in high-throughput ab initio databases*, Comput. Mater. Sci. **182**, 109747 (2020).

[28] G. R. Schleder, C. M. Acosta, and A. Fazzio, *Exploring Two-Dimensional Materials Thermodynamic Stability via Machine Learning*, ACS Appl. Mater. Interfaces **12**, 20149–20157 (2020).

[29] A. Puthirath Balan, S. Radhakrishnan, C. F. Woellner, S. K. Sinha, L. Deng, C. d. l. Reyes, B. M. Rao, M. Paulose, R. Neupane, A. Apte, V. Kochat, R. Vajtai, A. R. Harutyunyan, C.-W. Chu, G. Costin, D. S. Galvao, A. A. Martí, P. A. van Aken, O. K. Varghese, C. S. Tiwary, A. Malie Madom Ramaswamy Iyer, and P. M. Ajayan, *Exfoliation of a non-van der Waals material from iron ore hematite*, Nat. Nanotechnol. **13**, 602–609 (2018).

[30] A. Puthirath Balan, S. Radhakrishnan, R. Kumar, R. Neupane, S. K. Sinha, L. Deng, C. A. de los Reyes, A. Apte, B. M. Rao, M. Paulose, R. Vajtai, C. W. Chu, G. Costin, A. A. Martí, O. K. Varghese, A. K. Singh, C. S. Tiwary, M. R. Anantharaman, and P. M. Ajayan, *A Non-van der Waals Two-Dimensional Material from Natural Titanium Mineral Ore Ilmenite*, Chem. Mater. **30**, 5923–5931 (2018).

[31] H. Kaur, R. Tian, A. Roy, M. McCrystall, D. V. Horvath, G. Lozano Onrubia, R. Smith, M. Ruether, A. Griffin, C. Backes, V. Nicolosi, and J. N. Coleman, *Production of Quasi-2D Platelets of Nonlayered Iron Pyrite (FeS₂) by Liquid-Phase Exfoliation for High Performance Battery Electrodes*, ACS Nano **14**, 13418–13432 (2020).

[32] A. B. Puthirath, A. P. Balan, E. F. Oliveira, L. Deng, R. Dahal, F. C. R. Hernandez, G. Gao, N. Chakingal, L. M. Sassi, T. Prasankumar, G. Costin, R. Vajtai, C. W. Chu, D. S. Galvao, R. R. Nair, and P. M. Ajayan, *Exfoliated Pyrite (FeS2)- a non-van der Waals 2D Ferromagnet*, arXiv:2010.03113 [cond-mat, physics:physics] (2020). ArXiv: 2010.03113.

[33] M. G. Moinuddin, S. Srinivasan, and S. K. Sharma, *Probing Ferrimagnetic Semiconductor with Enhanced Negative Magnetoresistance: 2D Chromium Sulfide*, Adv. Electron. Mater. **n/a**, 2001116 (2021).

[34] L. Hu, L. Cao, L. Li, J. Duan, X. Liao, F. Long, J. Zhou, Y. Xiao, Y.-J. Zeng, and S. Zhou, *Two-dimensional magneto-photoconductivity in non-van der Waals manganese selenide*, Mater. Horiz. **8**, 1286–1296 (2021).

[35] A. Yousaf, M. S. Gilliam, S. L. Y. Chang, M. Augustin, Y. Guo, F. Tahir, M. Wang, A. Schwindt, X. S. Chu, D. O. Li, S. Kale, A. Debnath, Y. Liu, M. D. Green, E. J. G. Santos, A. A. Green, and Q. H. Wang, *Exfoliation of Quasi-Two-Dimensional Nanosheets of Metal Diborides*, J. Phys. Chem. C **125**, 6787–6799 (2021).

[36] Y. Guo, A. Gupta, M. S. Gilliam, A. Debnath, A. Yousaf, S. Saha, M. D. Levin, A. A. Green, A. K. Singh, and Q. H. Wang, *Exfoliation of boron carbide into ultrathin nanosheets*, Nanoscale **13**, 1652–1662 (2021).

[37] C. Gibaja, D. Rodríguez-San-Miguel, W. S. Paz, I. Torres, E. Salagre, P. Segovia, E. G. Michel, M. Assebban, P. Ares, D. Hernández-Maldonado, Q. Ramasse, G. Abellán, J. Gómez-Herrero, M. Varela, J. J. Palacios, and F. Zamora, *Exfoliation of Alpha-Germanium: A Covalent Diamond-Like Structure*, Adv. Mater. **33**, 2006826 (2021).

[38] Y. Wei, M. Ghorbani-Asl, and A. V. Krasheninnikov, *Tailoring the Electronic and Magnetic Properties of Hematene by Surface Passivation: Insights from First-Principles Calculations*, J. Phys. Chem. C **124**, 22784–22792 (2020).

[39] B. Y. Zhang, K. Xu, Q. Yao, A. Jannat, G. Ren, M. R. Field, X. Wen, C. Zhou, A. Zavabeti, and J. Z. Ou, *Hexagonal metal oxide monolayers derived from the metal–gas interface*, Nat. Mater. pp. 1–6 (2021).

[40] Y. Wei, C. Liu, T. Wang, Y. Zhang, C. Qi, H. Li, G. Ma, Y. Liu, S. Dong, and M. Huo, *Electronegativity regulation on opt-electronic properties of non van der Waals two-dimensional material: Ga₂O₃*, Comput. Mater. Sci. **179**, 109692 (2020).

[41] C. Jin and L. Kou, *Two-dimensional non-van der Waals magnetic layers: functional materials for potential device applications*, J. Phys. D: Appl. Phys. **54**, 413001 (2021).

[42] A. C. M. Padilha, M. Soares, E. R. Leite, and A. Fazzio, *Theoretical and Experimental Investigation of 2D Hematite*, J. Phys. Chem. C **123**, 16359–16365 (2019).

[43] R. I. Gonzalez, J. Mella, P. Díaz, S. Allende, E. E. Vogel, C. Cardenas, and F. Munoz, *Hematene: a 2D magnetic material in van der Waals or non-van der Waals heterostructures*, 2D Mater. **6**, 045002 (2019).

[44] A. Bandyopadhyay, N. C. Frey, D. Jariwala, and V. B. Shenoy, *Engineering Magnetic Phases in Two-Dimensional Non-van der Waals Transition-Metal Oxides*, Nano Lett. **19**, 7793–7800 (2019).

[45] S. Chahal, S. M. Kauzlarich, and P. Kumar, *Microwave Synthesis of Hematene and Other Two-Dimensional Oxides*, ACS Materials Lett. **3**, 631–640 (2021).

[46] J. Mohapatra, A. Ramos, J. Elkins, J. Beatty, M. Xing, D. Singh, E. C. La Plante, and J. Ping Liu, *Ferromagnetism in 2D α-Fe₂O₃ nanosheets*, Appl. Phys. Lett. **118**, 183102 (2021).

[47] K. Momma and F. Izumi, *VESTA 3 for three-dimensional visualization of crystal, volumetric and morphology data*, J. Appl. Crystallogr. **44**, 1272–1276 (2011).

[48] S. Liu, L. Xie, H. Qian, G. Liu, H. Zhong, and H. Zeng, *Facile preparation of novel and active 2D nanosheets from non-layered and traditionally non-exfoliable earth-abundant materials*, J. Mater. Chem. A **7**, 15411–15419 (2019).

[49] S. Curtarolo, W. Setyawan, S. Wang, J. Xue, K. Yang, R. H. Taylor, L. J. Nelson, G. L. W. Hart, S. Sanvito, M. Buongiorno Nardelli, N. Mingo, and O. Levy,

*AFLOWLIB.ORG: A distributed materials properties repository from high-throughput ab initio calculations*, Comput. Mater. Sci. **58**, 227–235 (2012).

[50] D. Hicks, C. Oses, E. Gossett, G. Gomez, R. H. Taylor, C. Toher, M. J. Mehl, O. Levy, and S. Curtarolo, *AFLOW-SYM: platform for the complete, automatic and self-consistent symmetry analysis of crystals*, Acta Crystallogr. Sect. A **74**, 184–203 (2018).

[51] D. Hicks, C. Toher, D. C. Ford, F. Rose, C. De Santo, O. Levy, M. J. Mehl, and S. Curtarolo, *AFLOW-XtalFinder: a reliable choice to identify crystalline prototypes*, npj Comput. Mater. **7**, 30 (2021).

[52] Structural data are retrieved via the AFLOW APIs [79, 80] and web interfaces [49] as well as the library of crystallographic prototypes [81].

[53] C. Backes, D. Campi, B. M. Szydlowska, K. Synnatschke, E. Ojala, F. Rashvand, A. Harvey, A. Griffin, Z. Sofer, N. Marzari, J. N. Coleman, and D. D. O'Regan, *Equipartition of Energy Defines the Size–Thickness Relationship in Liquid-Exfoliated Nanosheets*, ACS Nano **13**, 7050–7061 (2019).

[54] R. Zacharia, H. Ulbricht, and T. Hertel, *Interlayer cohesive energy of graphite from thermal desorption of polyaromatic hydrocarbons*, Phys. Rev. B **69**, 155406 (2004).

[55] T. Björkman, A. Gulans, A. V. Krasheninnikov, and R. M. Nieminen, *van der Waals Bonding in Layered Compounds from Advanced Density-Functional First-Principles Calculations*, Phys. Rev. Lett. **108**, 235502 (2012).

[56] J. Sun, R. C. Remsing, Y. Zhang, Z. Sun, A. Ruzsinszky, H. Peng, Z. Yang, A. Paul, U. Waghmare, X. Wu, M. L. Klein, and J. P. Perdew, *Accurate first-principles structures and energies of diversely bonded systems from an efficient density functional*, Nat. Chem. **8**, 9 (2016).

[57] B. Mohanty, Y. Wei, M. Ghorbani-Asl, A. V. Krasheninnikov, P. Rajput, and B. K. Jena, *Revealing the defect-dominated oxygen evolution activity of hematene*, J. Mater. Chem. A **8**, 6709–6716 (2020).

[58] S. R. Bahn and K. W. Jacobsen, *An object-oriented scripting interface to a legacy electronic structure code*, Comput. Sci. Eng. **4**, 56–66 (2002).

[59] A. H. Larsen, J. J. Mortensen, J. Blomqvist, I. E. Castelli, R. Christensen, M. Dułak, J. Friis, M. N. Groves, B. Hammer, C. Hargus, E. D. Hermes, P. C. Jennings, P. B. Jensen, J. Kermode, J. R. Kitchin, E. L. Kolsbjerg, J. Kubal, K. Kaasbjerg, S. Lysgaard, J. B. Maronsson, T. Maxson, T. Olsen, L. Pastewka, A. Peterson, C. Rostgaard, J. Schiøtz, O. Schütt, M. Strange, K. S. Thygesen, T. Vegge, L. Vilhelmsen, M. Walter, Z. Zeng, and K. W. Jacobsen, *The atomic simulation environment—a Python library for working with atoms*, Journal of Physics: Condensed Matter **29**, 273002 (2017).

[60] S. Blundell, *Magnetism in Condensed Matter* (Oxford University Press, Oxford, 2009).

[61] C. E. Calderon, J. J. Plata, C. Toher, C. Oses, O. Levy, M. Fornari, A. Natan, M. J. Mehl, G. L. W. Hart, M. Buongiorno Nardelli, and S. Curtarolo, *The AFLOW standard for high-throughput materials science calculations*, Comput. Mater. Sci. **108 Part A**, 233–238 (2015).

[62] R. Friedrich, D. Usanmaz, C. Oses, A. Supka, M. Fornari, M. Buongiorno Nardelli, C. Toher, and S. Curtarolo, *Coordination corrected ab initio formation enthalpies*, npj Comput. Mater. **5**, 59 (2019).

[63] R. Friedrich, M. Esters, C. Oses, S. Ki, M. J. Brenner, D. Hicks, M. J. Mehl, C. Toher, and S. Curtarolo, *Automated coordination corrected enthalpies with AFLOW-CCE*, Phys. Rev. Mater. **5**, 043803 (2021).

[64] V. Heß, R. Friedrich, F. Matthes, V. Caciuc, N. Atodiresei, D. E. Bürgler, Stefan Blügel, and C. M. Schneider, *Magnetic subunits within a single molecule–surface hybrid*, New J. Phys. **19**, 053016 (2017).

[65] W. Kohn and L. J. Sham, *Self-consistent equations including exchange and correlation effects*, Phys. Rev. **140**, A1133 (1965).

[66] U. von Barth and L. Hedin, *A local exchange-correlation potential for the spin polarized case: I*, J. Phys. C: Solid State Phys. **5**, 1629 (1972).

[67] J. P. Perdew, K. Burke, and M. Ernzerhof, *Generalized Gradient Approximation Made Simple*, Phys. Rev. Lett. **77**, 3865–3868 (1996).

[68] J. Sun, A. Ruzsinszky, and J. P. Perdew, *Strongly Constrained and Appropriately Normed Semilocal Density Functional*, Phys. Rev. Lett. **115**, 036402 (2015).

[69] S. L. Dudarev, G. A. Botton, S. Y. Savrasov, C. J. Humphreys, and A. P. Sutton, *Electron-energy-loss spectra and the structural stability of nickel oxide: An LSDA+U study*, Phys. Rev. B **57**, 1505–1509 (1998).

[70] A. I. Liechtenstein, V. I. Anisimov, and J. Zaanen, *Density-functional theory and strong interactions: Orbital ordering in Mott-Hubbard insulators*, Phys. Rev. B **52**, R5467–R5470 (1995).

[71] V. I. Anisimov, J. Zaanen, and O. K. Andersen, *Band theory and Mott insulators: Hubbard U instead of Stoner I*, Phys. Rev. B **44**, 943–954 (1991).

[72] O. Levy, R. V. Chepulskii, G. L. W. Hart, and S. Curtarolo, *The New Face of Rhodium Alloys: Revealing Ordered Structures from First Principles*, J. Am. Chem. Soc. **132**, 833–837 (2010).

[73] S. Curtarolo, W. Setyawan, G. L. W. Hart, M. Jahnátek, R. V. Chepulskii, R. H. Taylor, S. Wang, J. Xue, K. Yang, O. Levy, M. J. Mehl, H. T. Stokes, D. O. Demchenko, and D. Morgan, *AFLOW: An automatic framework for high-throughput materials discovery*, Comput. Mater. Sci. **58**, 218–226 (2012).

[74] G. Kresse and J. Hafner, *Ab initio molecular dynamics for liquid metals*, Phys. Rev. B **47**, 558–561 (1993).

[75] G. Kresse and J. Furthmüller, *Efficient iterative schemes for ab initio total-energy calculations using a plane-wave basis set*, Phys. Rev. B **54**, 11169–11186 (1996).

[76] G. Kresse and J. Furthmüller, *Efficiency of ab-initio total energy calculations for metals and semiconductors using a plane-wave basis set*, Comput. Mater. Sci. **6**, 15–50 (1996).

[77] G. Kresse and D. Joubert, *From ultrasoft pseudopotentials to the projector augmented-wave method*, Phys. Rev. B **59**, 1758–1775 (1999).

[78] R. V. Chepulskii and S. Curtarolo, *First principles study of Ag, Au, and Cu surface segregation in FePt-L1₀*, Appl. Phys. Lett. **97**, 221908 (2010).

[79] R. H. Taylor, F. Rose, C. Toher, O. Levy, K. Yang, M. Buongiorno Nardelli, and S. Curtarolo, *A RESTful API for exchanging materials data in the AFLOWLIB.org consortium*, Comput. Mater. Sci. **93**, 178–192 (2014).

[80] F. Rose, C. Toher, E. Gossett, C. Oses, M. Buongiorno Nardelli, M. Fornari, and S. Curtarolo, *AFLUX: The LUX materials search API for the AFLOW data repositories*, Comput. Mater. Sci. **137**, 362–370 (2017).

[81] M. J. Mehl, D. Hicks, C. Toher, O. Levy, R. M. Hanson, G. L. W. Hart, and S. Curtarolo, *The AFLOW Library of*

*Crystallographic Prototypes: Part 1*, Comput. Mater. Sci. **136**, S1–S828 (2017).

# Data-driven quest for two-dimensional non-van der Waals materials
## Supporting Information

Rico Friedrich,[1, 2, *] Mahdi Ghorbani-Asl,[1] Stefano Curtarolo,[2, 3] and Arkady V. Krasheninnikov[1, 4]

[1]*Institute of Ion Beam Physics and Materials Research,*
*Helmholtz-Zentrum Dresden-Rossendorf, 01328 Dresden, Germany*
[2]*Center for Autonomous Materials Design, Duke University, Durham, North Carolina 27708, USA*
[3]*Materials Science, Electrical Engineering, and Physics, Duke University, Durham NC, 27708, USA*
[4]*Department of Applied Physics, Aalto University, Aalto 00076, Finland*
(Dated: September 30, 2021)

## I. Comparing inter-facet binding energies from different functionals

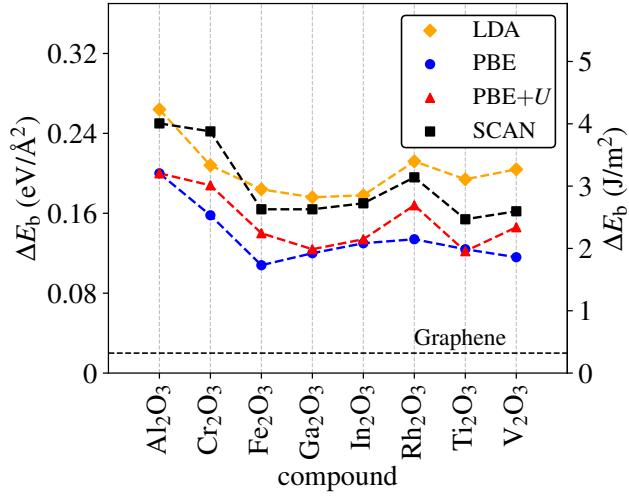

FIG. S1. **Inter-facet binding energies for binaries.** Binding energies of investigated binary systems for different functionals. Note that in case of $Al_2O_3$ PBE+$U$ reduces to PBE. As a reference, the inter-layer binding energy of graphene [1, 2] is indicated by the dashed horizontal black line. The dashed lines connecting the data points are visual guides.

In Fig. S1, the calculated binding energies of the binary systems for four different functionals are summarized. All approaches show similar trends and sizes of the computed values with LDA mostly providing estimates on the upper while PBE yields values on the lower end. The biggest absolute deviation between the functionals of 44 meV/$Å^2$ is observed for $V_2O_3$ between LDA and PBE. It can thus be concluded that the calculation of binding energies of these systems as energy differences between chemically similar bulk and 2D systems is rather insensitive to the choice of the functional. It can be performed efficiently within standard approximations.

## II. Comparing inter-facet binding energies from relaxing different degrees of freedom

After obtaining the 2D facets of the binaries from the bulk parent structures, we considered three different approaches for optimizing the structural degrees of freedom: (i) performing only a static self-consistent electronic calculation, (ii) also relaxing the ionic positions in the cell and (iii) also relaxing the (in-plane) cell parameters. The first approach has been employed in the large scale screening of vdW 2D materials [2] since for these systems the weak inter-layer bonding does typically not give rise to strong relaxations. For the latter approach, the VASP option to relax the cell shape

---

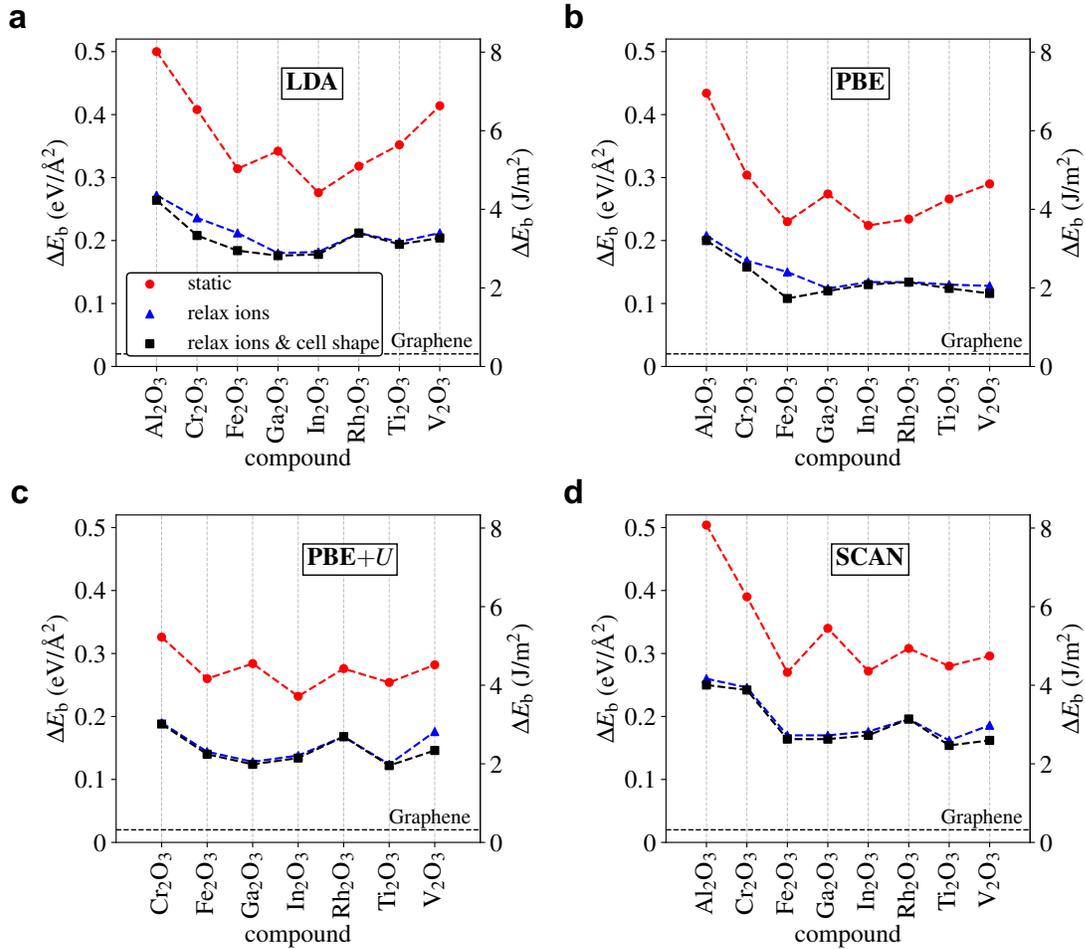

FIG. S2. **Binding energies from relaxing different degrees of freedom.** Comparison of binding energies calculated from only a static electronic calculation, a relaxation of the ionic positions as well as also relaxing the cell shape of the 2D systems for LDA (**a**), PBE (**b**), PBE+$U$ (**c**), and SCAN (**d**). For PBE+$U$, the results for $Al_2O_3$ are not plotted as they are the same as for PBE. The dashed lines connecting the data points are visual guides.

was chosen to allow the in-plane lattice parameters and angles to relax while keeping the overall volume, including the large vacuum distances to the periodic images, constant. The results are visualized in Fig. S2 for all four considered functionals. In each case, the binding energies are reduced by about a factor of two upon including ionic relaxations. Also relaxing the cell leads to only a minor additional energy gain of typically a few meV/Å². The largest energy gain upon cell relaxation is observed for $Fe_2O_3$ for PBE amounting to 21 meV/Å². In conclusion, while only performing a static calculation is not a reliable approximation for non-vdW 2D systems, the relaxation of ionic positions usually suffices to estimate their binding energy. For the investigation of the ternary systems, we have decided to consider only the full relaxation including cell parameters in this study.

## III. Comparing inter-facet binding energies for different terminations for ternary systems

Fig. S3 compares the binding energies of the ternary 2D systems with different cation terminations. In general, the difference between the energetically preferred termination compared to the less stable one is substantial amounting to a factor of two to five for both functionals. The SCAN binding energies are usually higher than the ones from PBE+$U$ for both terminations. A special case is $GeMnO_3$. PBE+$U$ yields the Mn-termination to be lower in energy while SCAN favors the Ge-termination. This qualitatively different behavior can be traced back to different oxidation numbers found by the approaches indicated by the charges and magnetic moments. While PBE+$U$ finds Mn and Ge to be in oxidation states +2 and +4, respectively, SCAN yields the opposite. This can be assigned to the well known self-interaction error that has been pointed out to persist for this functional [3]. That SCAN gives a lower $\Delta E_b$ than PBE+$U$ for the Bi

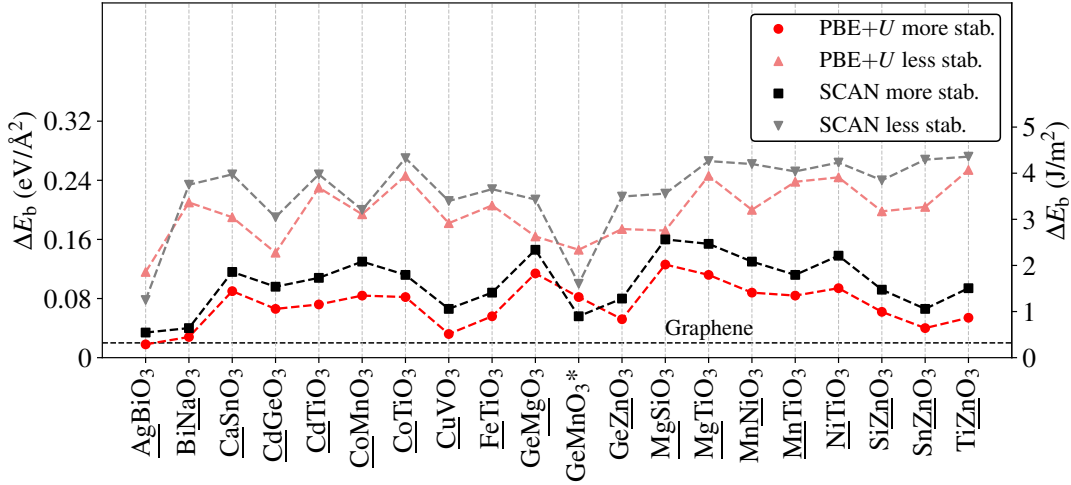

FIG. S3. **Binding energies for different terminations for ternary systems.** The terminating elements for the energetically more stable slabs are underlined. In case of GeMnO₃ (marked by "*"), PBE+$U$ favors Mn termination while for SCAN Ge termination is preferred. The dashed lines connecting the data points are visual guides.

terminated slabs for AgBiO₃ has likely a similar reason.

## IV. Change of structural parameters for SCAN

As indicated in Fig. S4, also for SCAN the expansion of the in-plane lattice parameters correlates well with the thickness reduction of the slabs upon exfoliation.

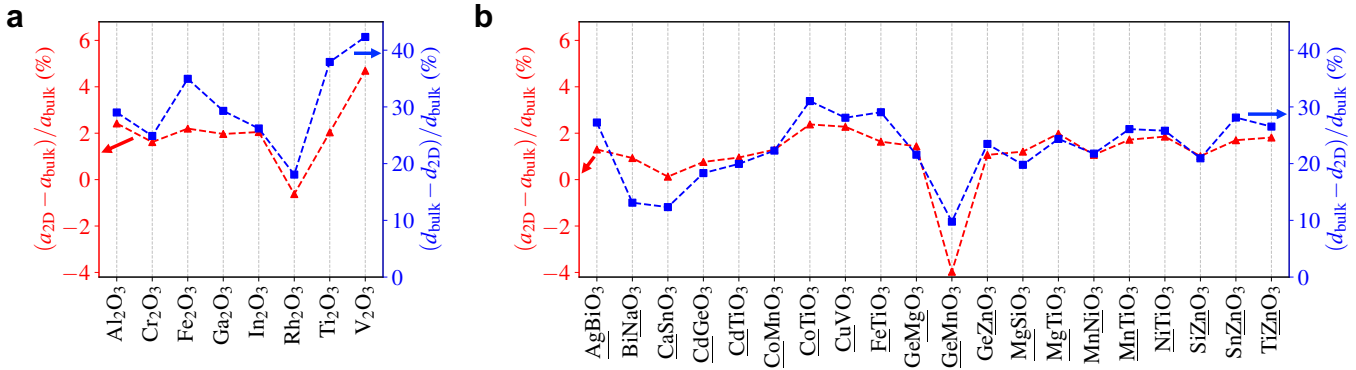

FIG. S4. **Modification of structural parameters.** Change of structural parameters for binary (**a**) and ternary (**b**) materials upon exfoliation as calculated with SCAN. The modification of the in-plane lattice parameter is indicated by the red curves (left $y$-axis) while the thickness change is given by the blue curves (right $y$-axis). The ternary data are for the energetically preferred slab termination. The dashed lines connecting the data points are visual guides.

## V. Band structures

The band structures and corresponding densities of states (DOSs) of the bulk and 2D systems as obtained from PBE+$U$ are presented in Figs. S5 to S32. A general observeration is that, as expected, the bands show less dispersion for the 2D systems compared to their bulk counterparts due to the missing out of plane interaction. For the 2D sheets of Ti₂O₃ and V₂O₃, the hexagonal symmetry of the cell was slightly broken during the relaxation such that the automatic analysis of AFLOW finds a different $k$-path as for the other systems. Therefore, in these cases only the DOS is presented. For AgBiO₃ and BiNaO₃ linear band crossings at the $K$-point are observed hinting at potentially interesting topological features.

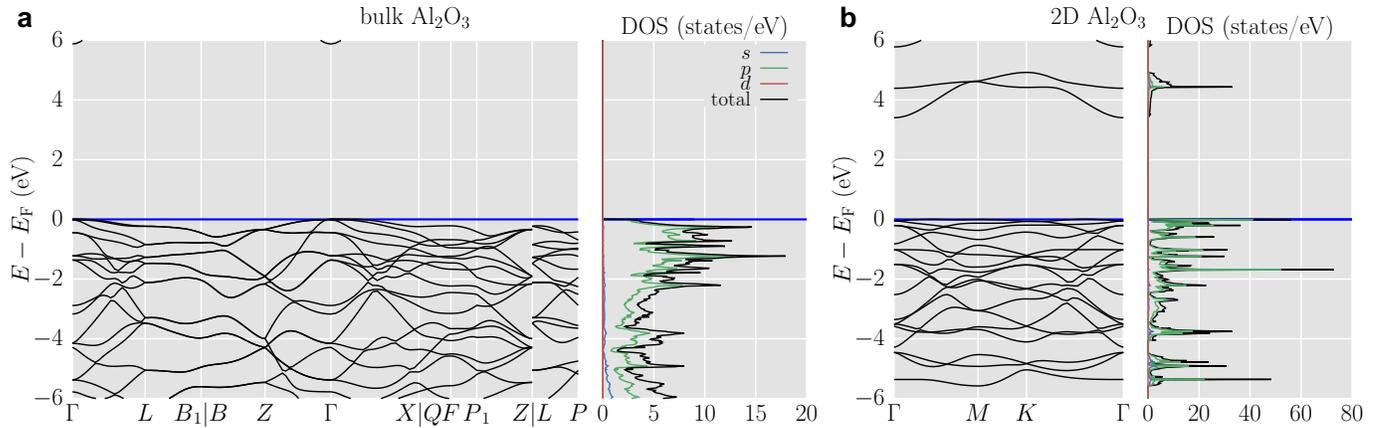

FIG. S5. Bandstructure and density of states for bulk (**a**) and 2D (**b**) Al$_2$O$_3$. The energies are aligned at the respective Fermi energy $E_F$.

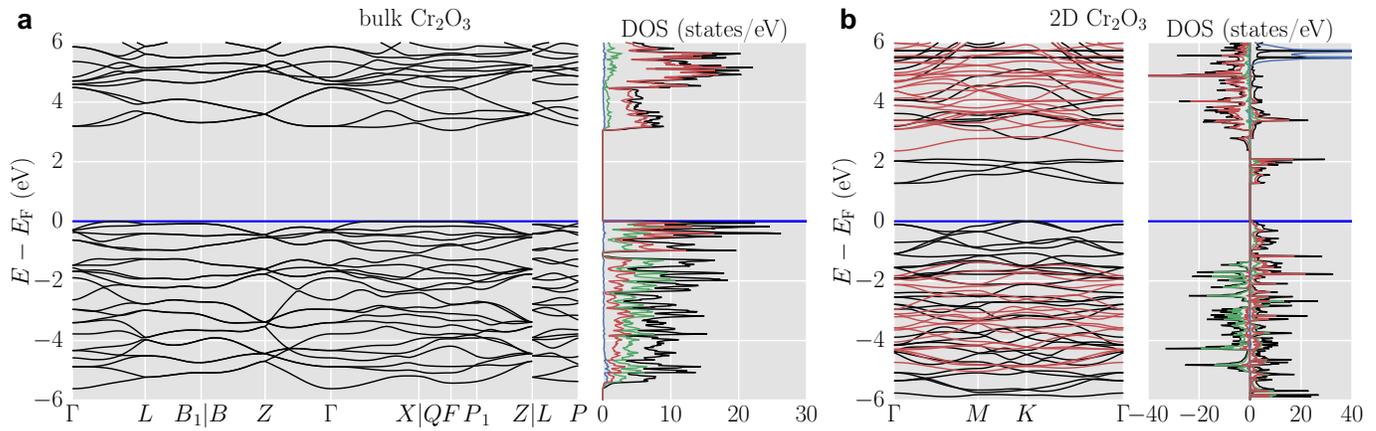

FIG. S6. Bandstructure and density of states for bulk (**a**) and 2D (**b**) Cr$_2$O$_3$. The energies are aligned at the respective Fermi energy $E_F$. For the spin polarized bandstructure, majority spin bands (positive DOS) are indicated in black while minority spin bands (negative DOS) are in red.

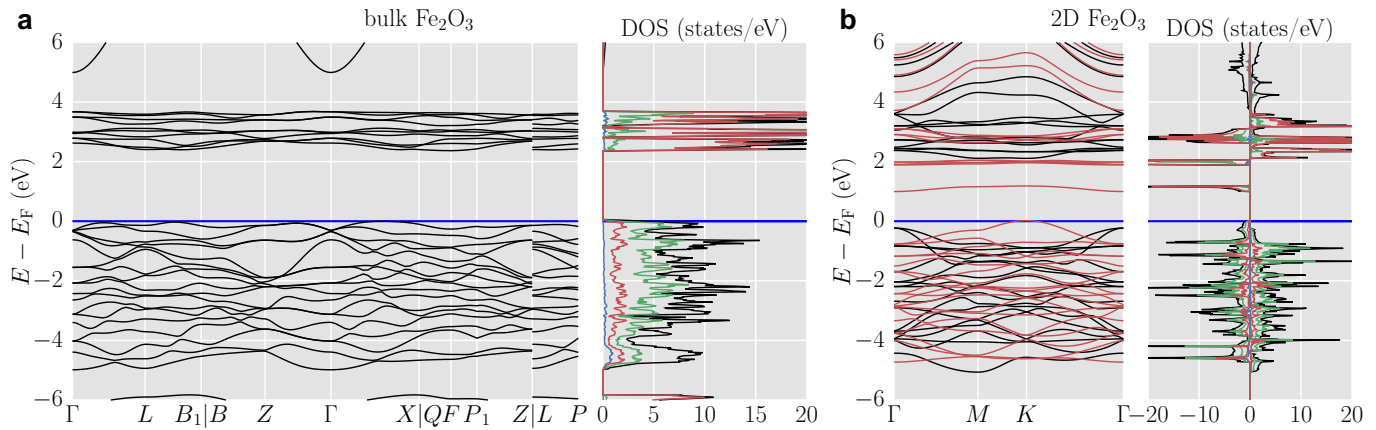

FIG. S7. Bandstructure and density of states for bulk (**a**) and 2D (**b**) Fe$_2$O$_3$. The energies are aligned at the respective Fermi energy $E_F$. For the spin polarized bandstructure, majority spin bands (positive DOS) are indicated in black while minority spin bands (negative DOS) are in red.

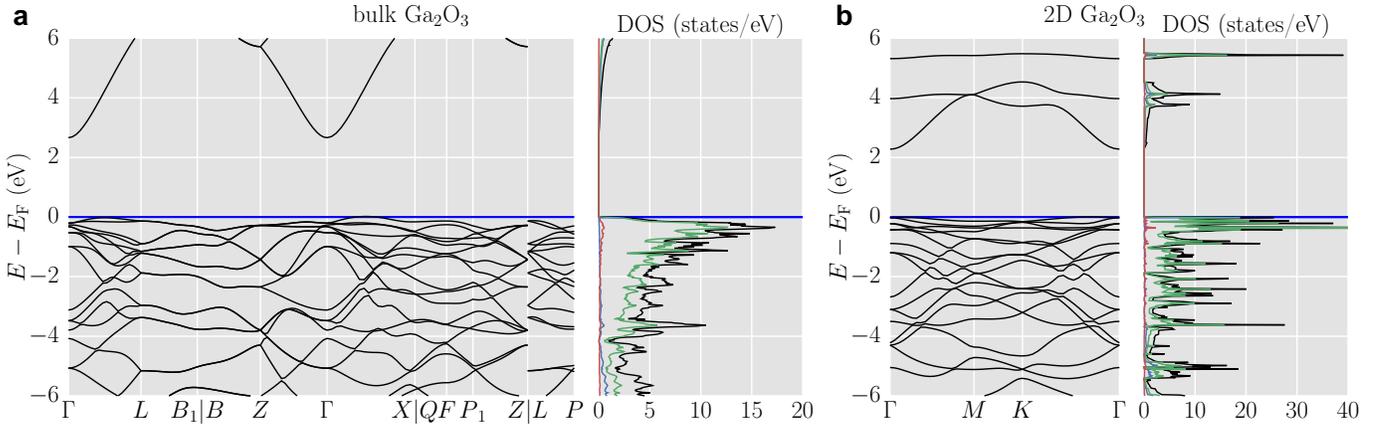

FIG. S8. Bandstructure and density of states for bulk (**a**) and 2D (**b**) $Ga_2O_3$. The energies are aligned at the respective Fermi energy $E_F$.

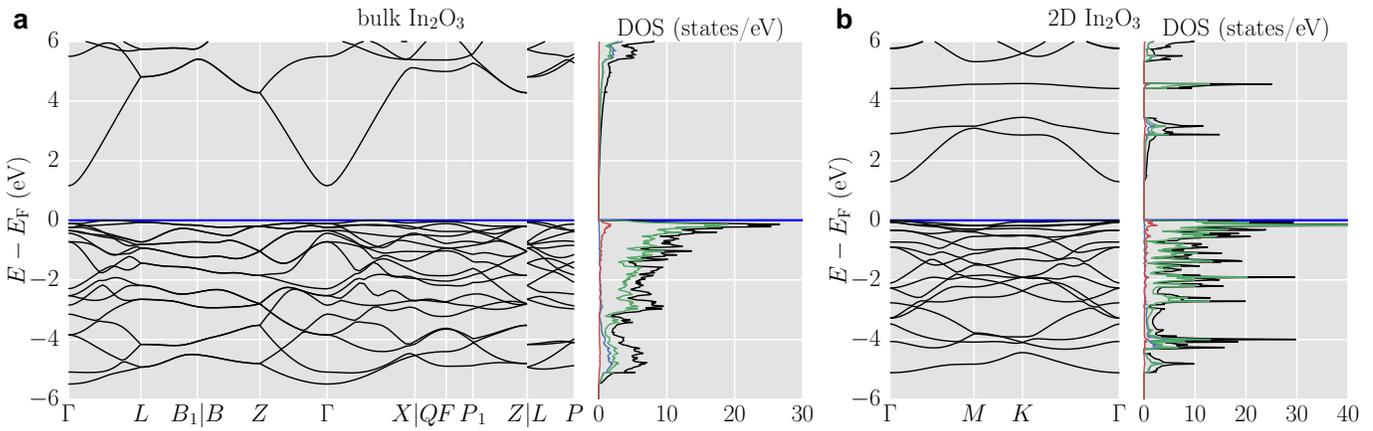

FIG. S9. Bandstructure and density of states for bulk (**a**) and 2D (**b**) $In_2O_3$. The energies are aligned at the respective Fermi energy $E_F$.

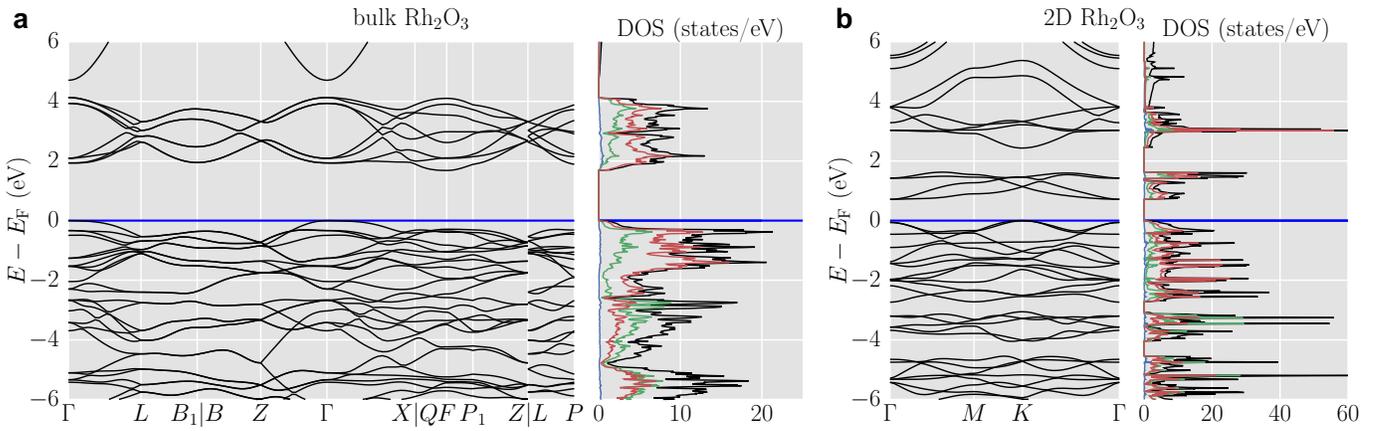

FIG. S10. Bandstructure and density of states for bulk (**a**) and 2D (**b**) $Rh_2O_3$. The energies are aligned at the respective Fermi energy $E_F$.

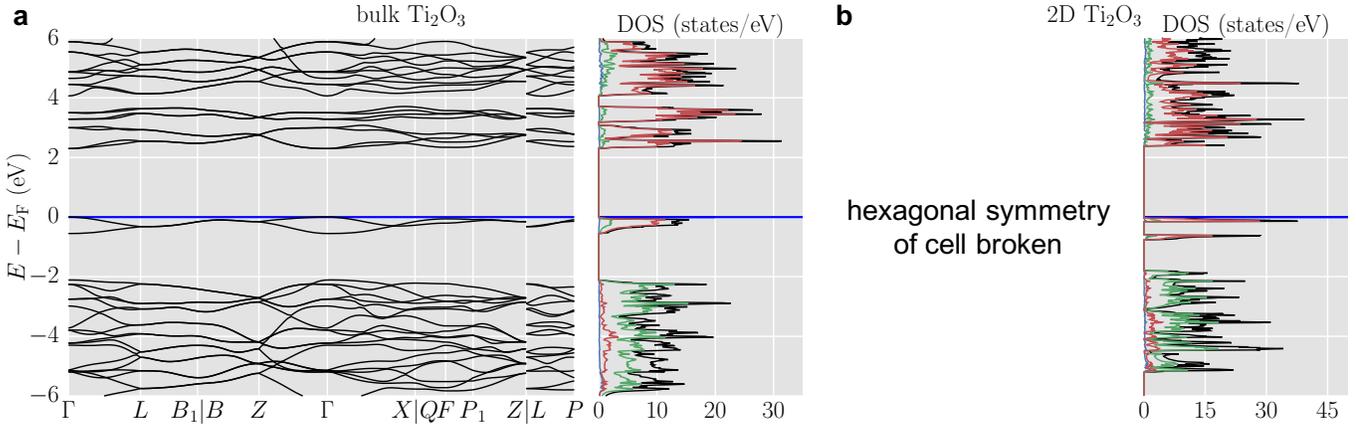

FIG. S11. Bandstructure and density of states for bulk (**a**) and 2D (**b**) Ti$_2$O$_3$. The energies are aligned at the respective Fermi energy $E_{\mathrm{F}}$.

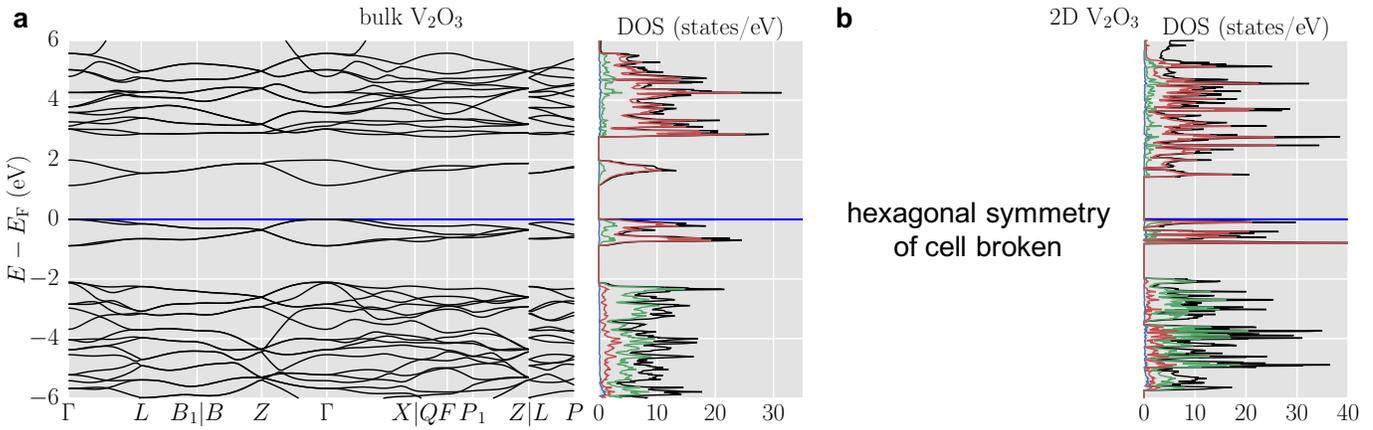

FIG. S12. Bandstructure and density of states for bulk (**a**) and 2D (**b**) V$_2$O$_3$. The energies are aligned at the respective Fermi energy $E_{\mathrm{F}}$.

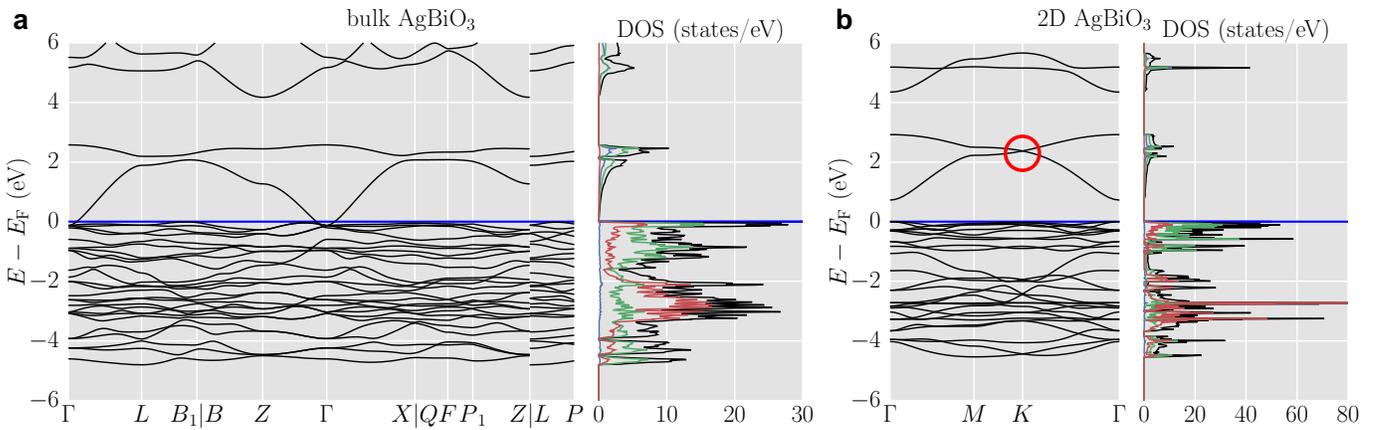

FIG. S13. Bandstructure and density of states for bulk (**a**) and 2D (**b**) AgBiO$_3$. The energies are aligned at the respective Fermi energy $E_{\mathrm{F}}$. A linear band crossing at the $K$-point is highlighted by the circle.

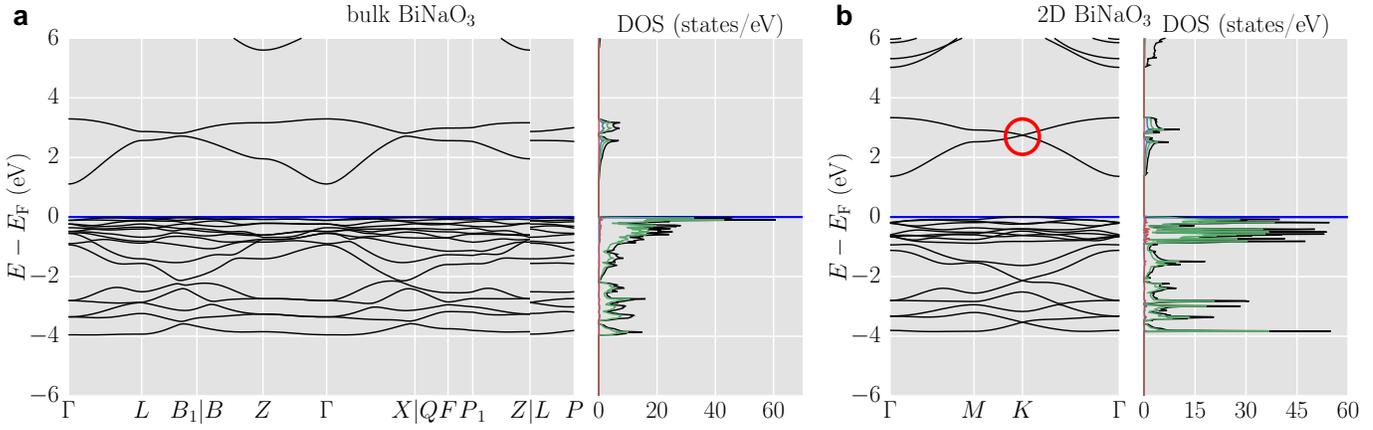

FIG. S14. Bandstructure and density of states for bulk (**a**) and 2D (**b**) BiNaO$_3$. The energies are aligned at the respective Fermi energy $E_F$. A linear band crossing at the $K$-point is highlighted by the circle.

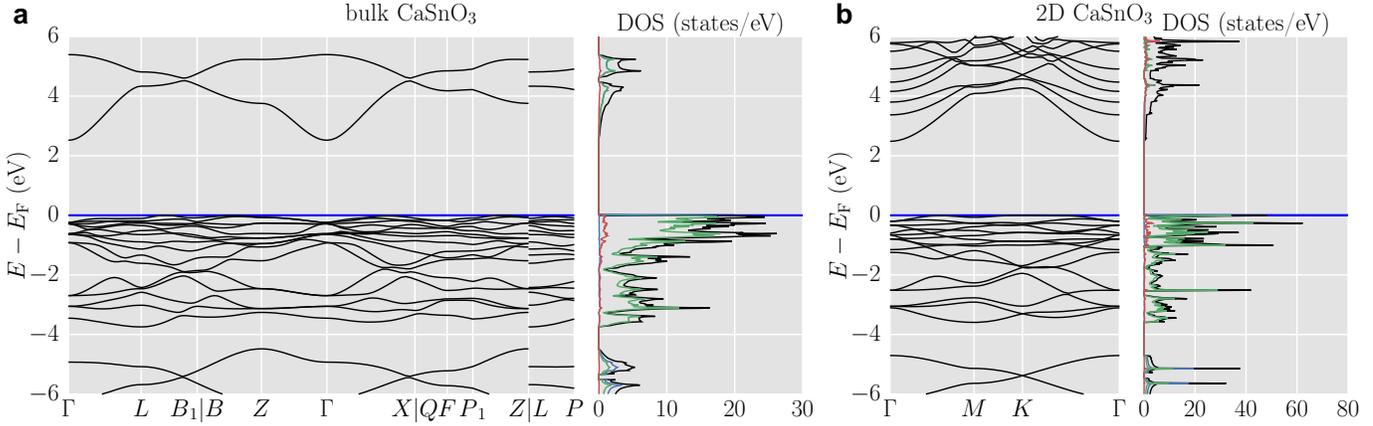

FIG. S15. Bandstructure and density of states for bulk (**a**) and 2D (**b**) CaSnO$_3$. The energies are aligned at the respective Fermi energy $E_F$.

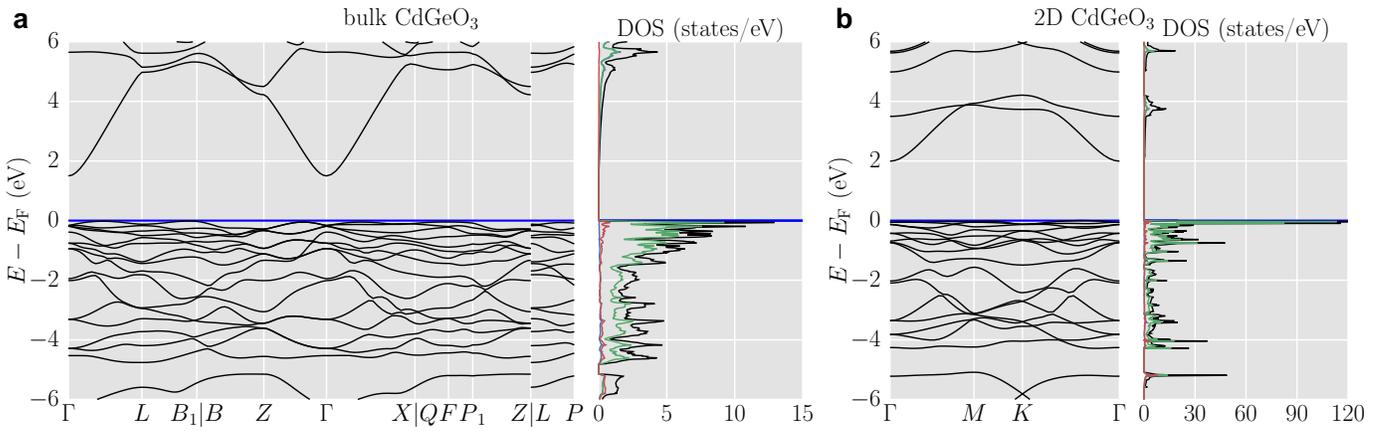

FIG. S16. Bandstructure and density of states for bulk (**a**) and 2D (**b**) CdGeO$_3$. The energies are aligned at the respective Fermi energy $E_F$.

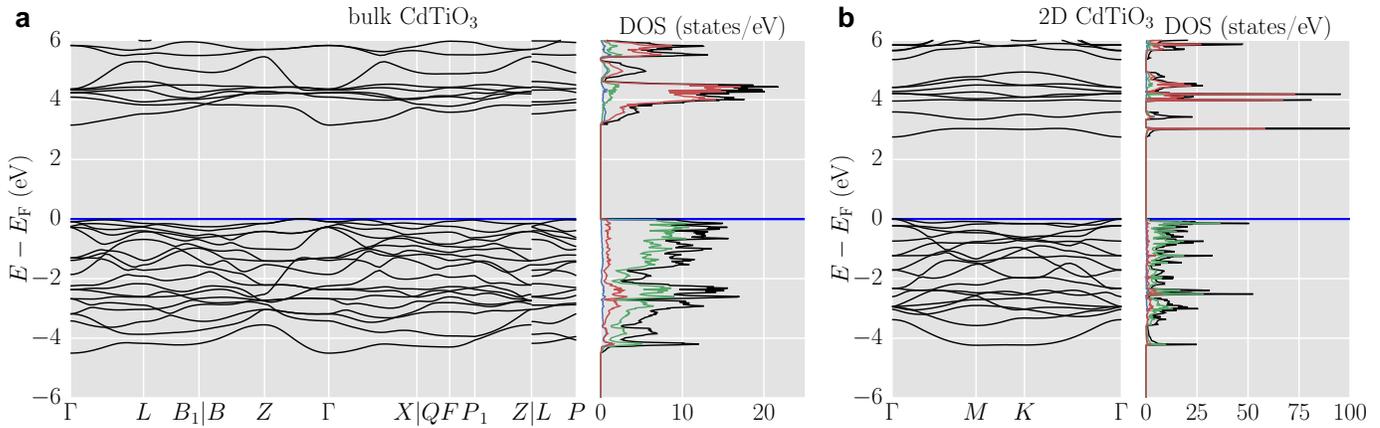

FIG. S17. Bandstructure and density of states for bulk (**a**) and 2D (**b**) CdTiO$_3$. The energies are aligned at the respective Fermi energy $E_F$.

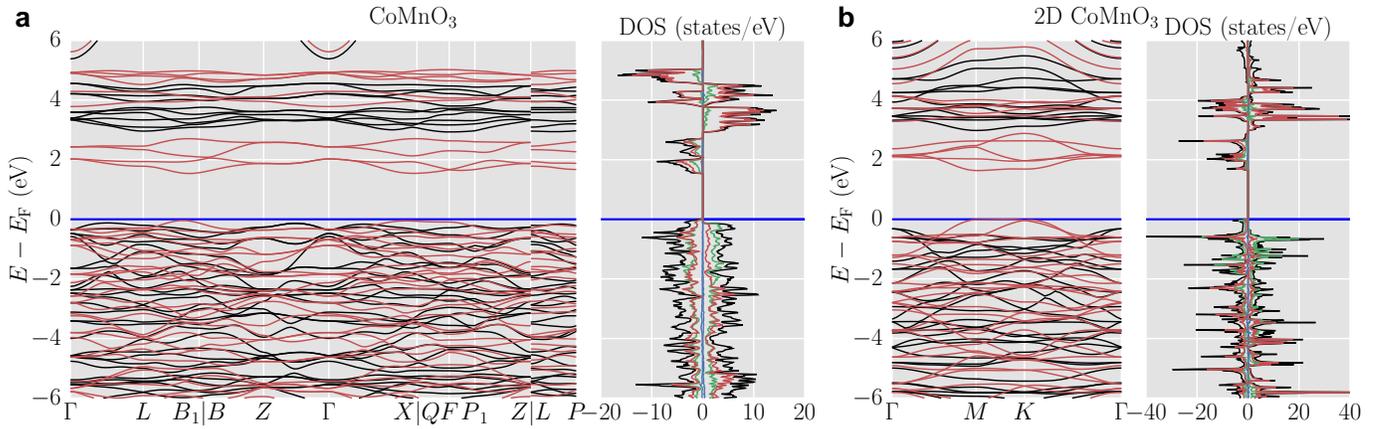

FIG. S18. Bandstructure and density of states for bulk (**a**) and 2D (**b**) CoMnO$_3$. The energies are aligned at the respective Fermi energy $E_F$. For the spin polarized bandstructure, majority spin bands (positive DOS) are indicated in black while minority spin bands (negative DOS) are in red.

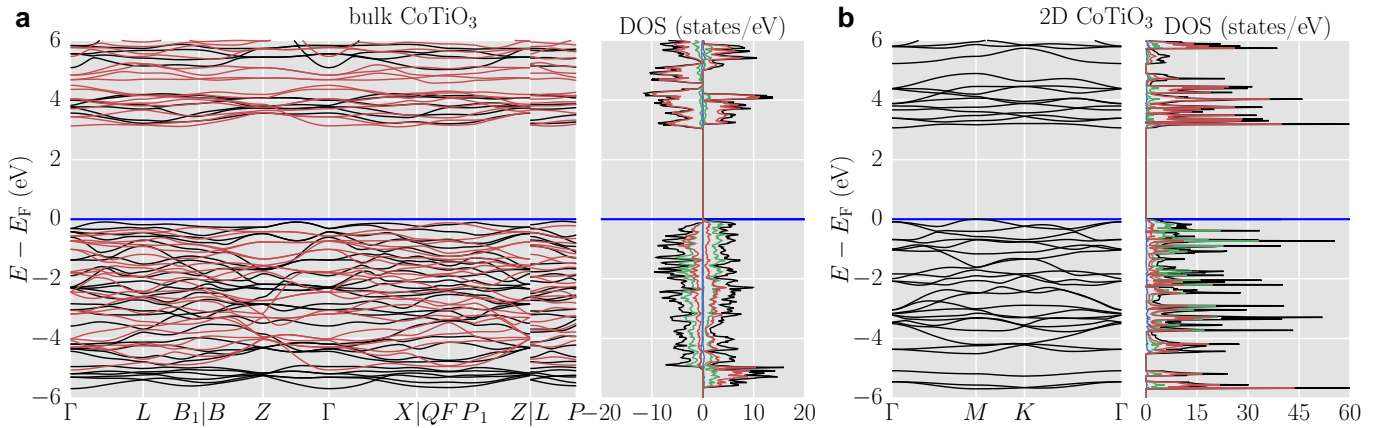

FIG. S19. Bandstructure and density of states for bulk (**a**) and 2D (**b**) CoTiO$_3$. The energies are aligned at the respective Fermi energy $E_F$. For the spin polarized bandstructure, majority spin bands (positive DOS) are indicated in black while minority spin bands (negative DOS) are in red.

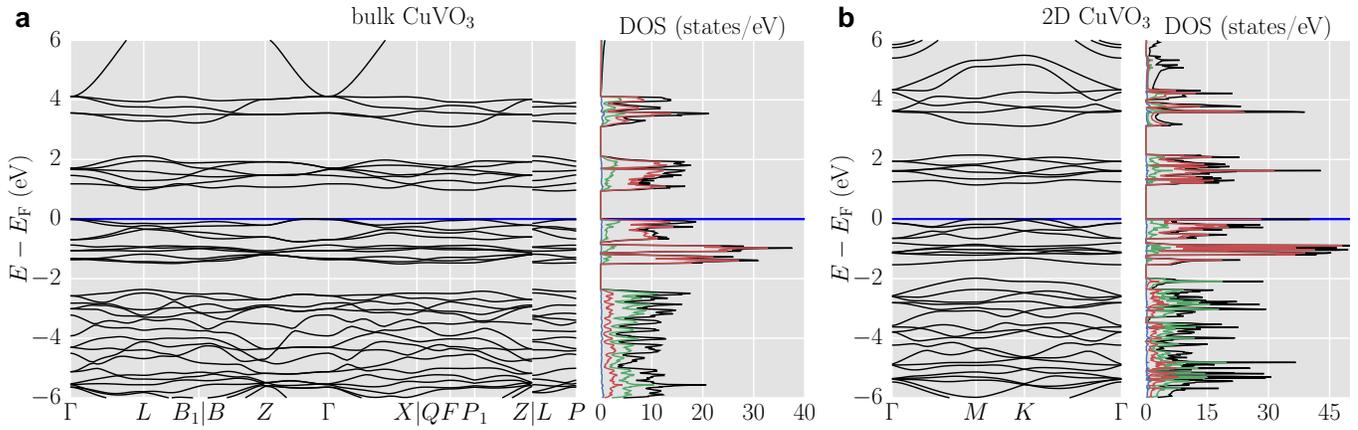

FIG. S20. Bandstructure and density of states for bulk (**a**) and 2D (**b**) CuVO$_3$. The energies are aligned at the respective Fermi energy $E_\mathrm{F}$.

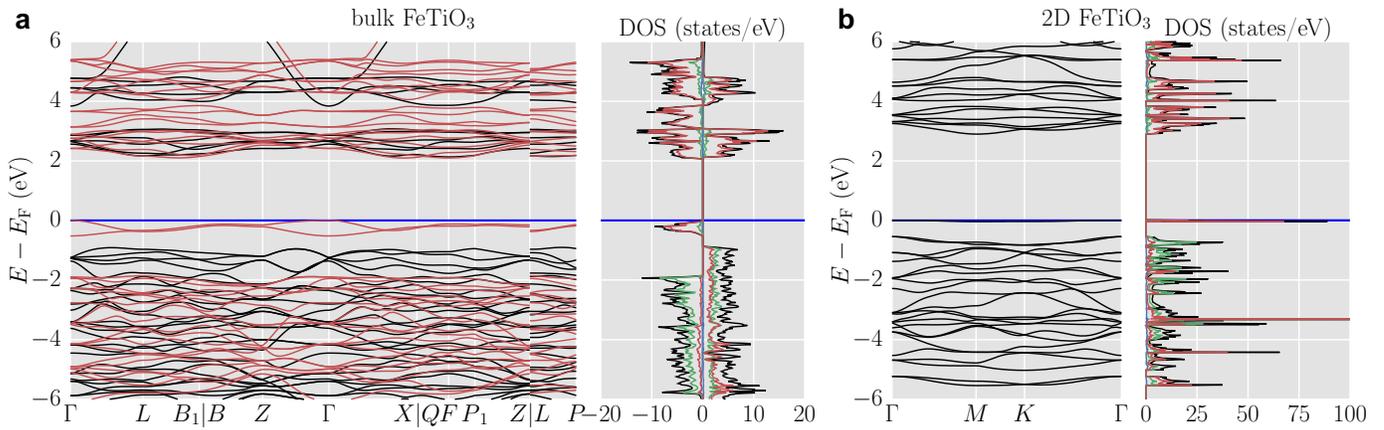

FIG. S21. Bandstructure and density of states for bulk (**a**) and 2D (**b**) FeTiO$_3$. The energies are aligned at the respective Fermi energy $E_\mathrm{F}$. For the spin polarized bandstructure, majority spin bands (positive DOS) are indicated in black while minority spin bands (negative DOS) are in red.

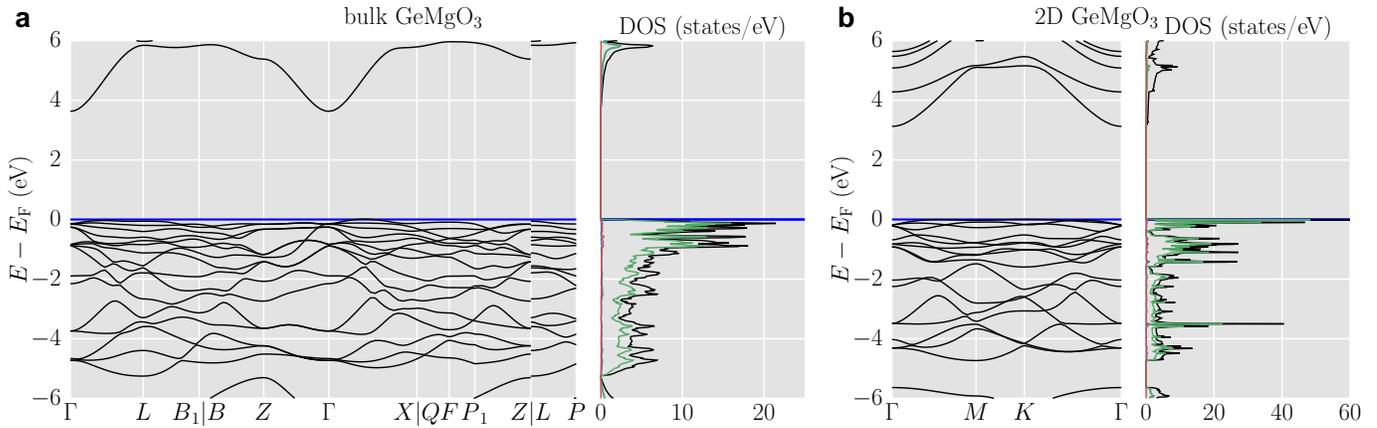

FIG. S22. Bandstructure and density of states for bulk (**a**) and 2D (**b**) GeMgO$_3$. The energies are aligned at the respective Fermi energy $E_\mathrm{F}$.

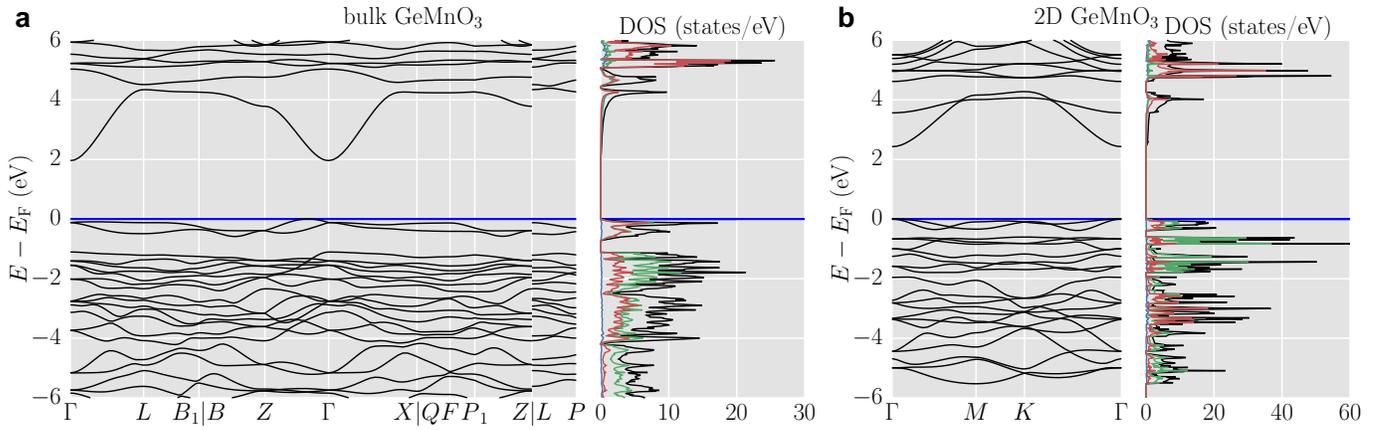

FIG. S23. Bandstructure and density of states for bulk (**a**) and 2D (**b**) GeMnO$_3$. The energies are aligned at the respective Fermi energy $E_{\mathrm{F}}$.

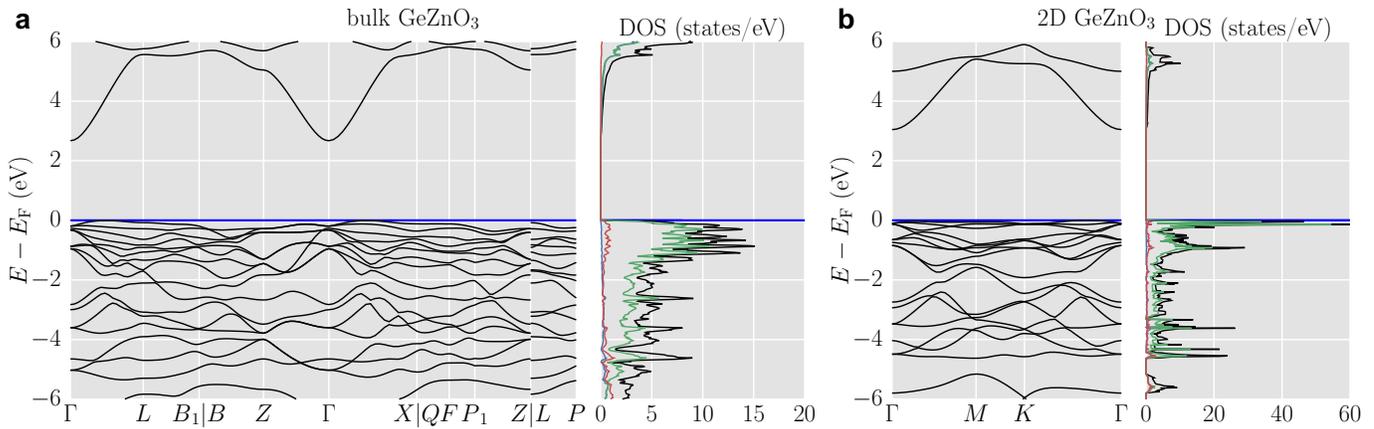

FIG. S24. Bandstructure and density of states for bulk (**a**) and 2D (**b**) GeZnO$_3$. The energies are aligned at the respective Fermi energy $E_{\mathrm{F}}$.

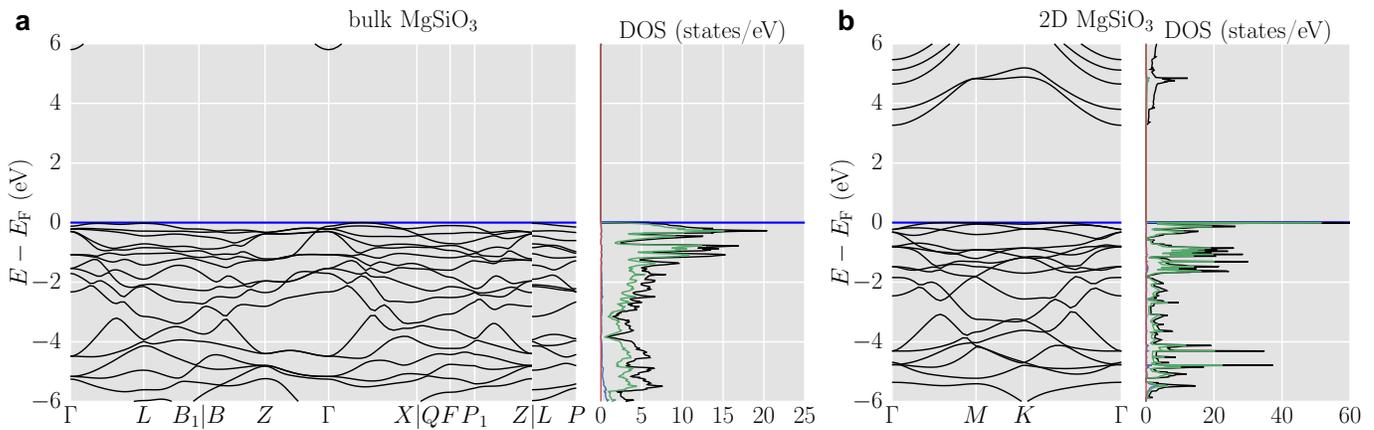

FIG. S25. Bandstructure and density of states for bulk (**a**) and 2D (**b**) MgSiO$_3$. The energies are aligned at the respective Fermi energy $E_{\mathrm{F}}$.

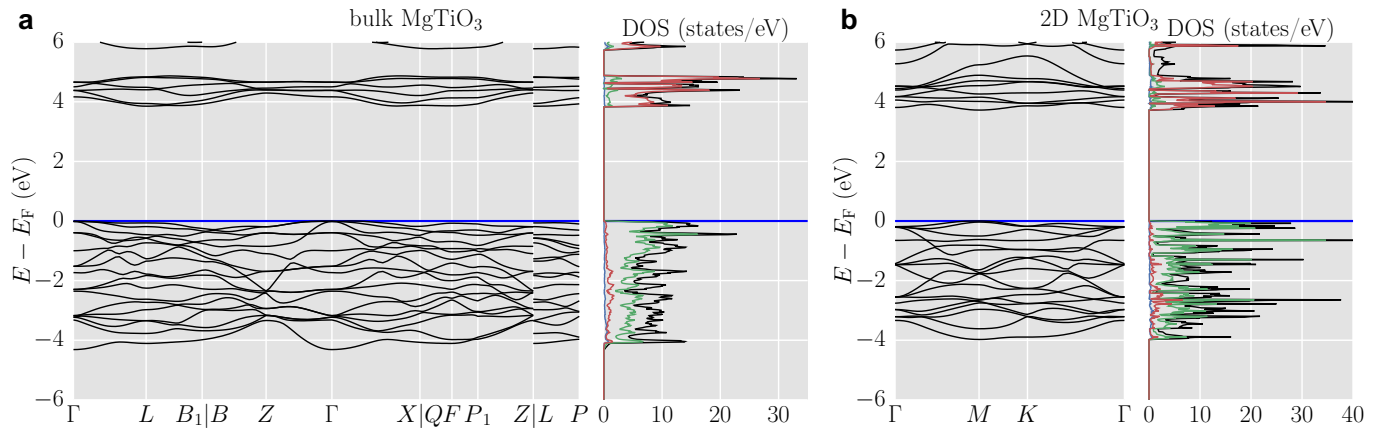

FIG. S26. Bandstructure and density of states for bulk (**a**) and 2D (**b**) MgTiO$_3$. The energies are aligned at the respective Fermi energy $E_\mathrm{F}$.

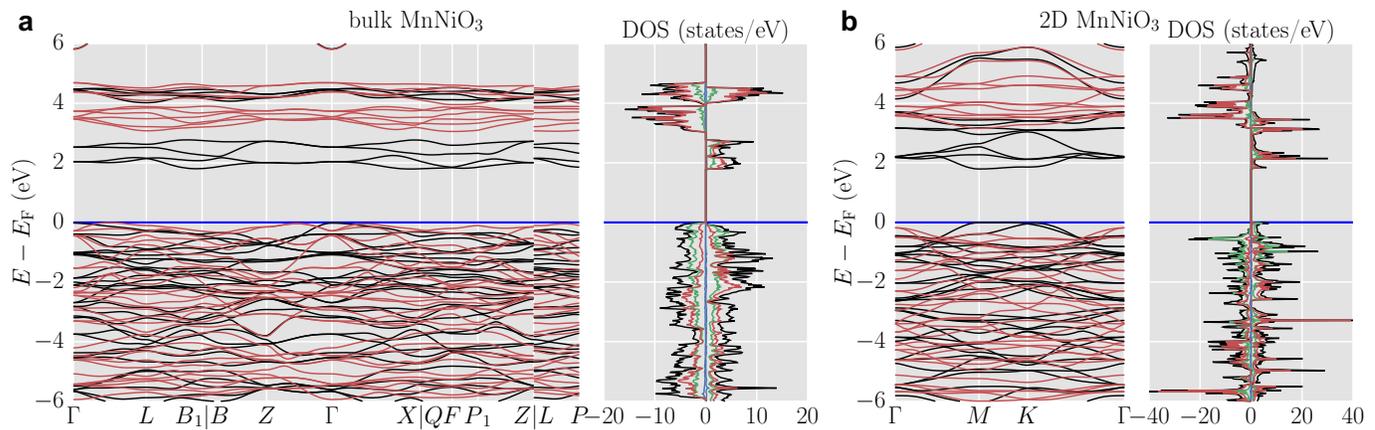

FIG. S27. Bandstructure and density of states for bulk (**a**) and 2D (**b**) MnNiO$_3$. The energies are aligned at the respective Fermi energy $E_\mathrm{F}$. For the spin polarized bandstructure, majority spin bands (positive DOS) are indicated in black while minority spin bands (negative DOS) are in red.

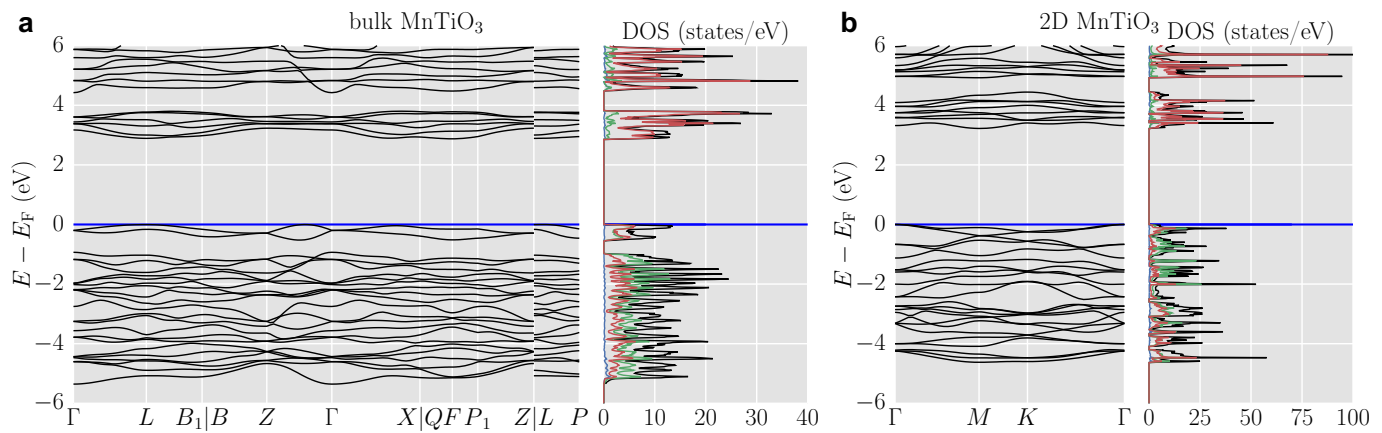

FIG. S28. Bandstructure and density of states for bulk (**a**) and 2D (**b**) MnTiO$_3$. The energies are aligned at the respective Fermi energy $E_\mathrm{F}$.

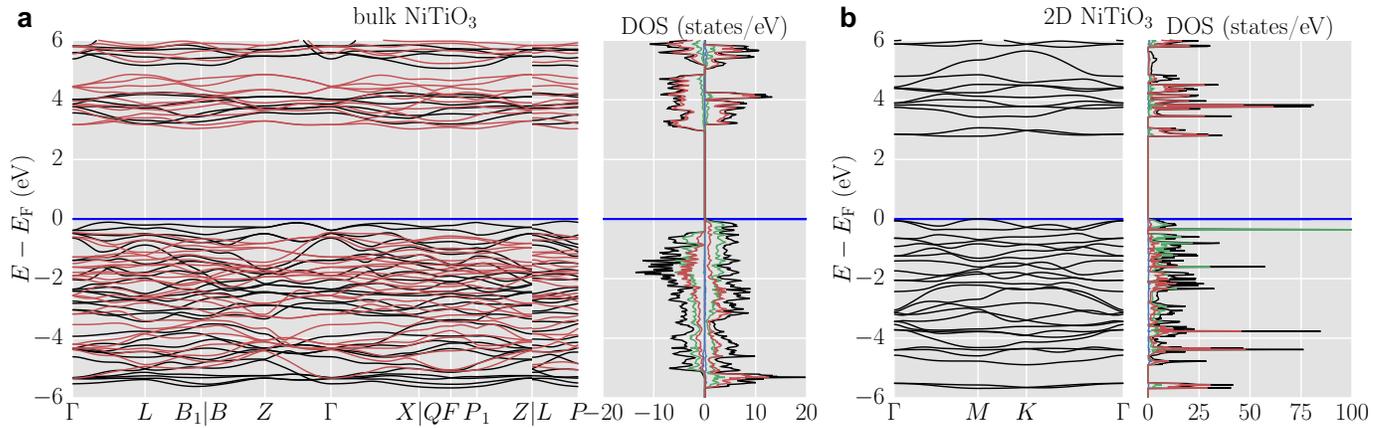

FIG. S29. Bandstructure and density of states for bulk (**a**) and 2D (**b**) NiTiO$_3$. The energies are aligned at the respective Fermi energy $E_F$. For the spin polarized bandstructure, majority spin bands (positive DOS) are indicated in black while minority spin bands (negative DOS) are in red.

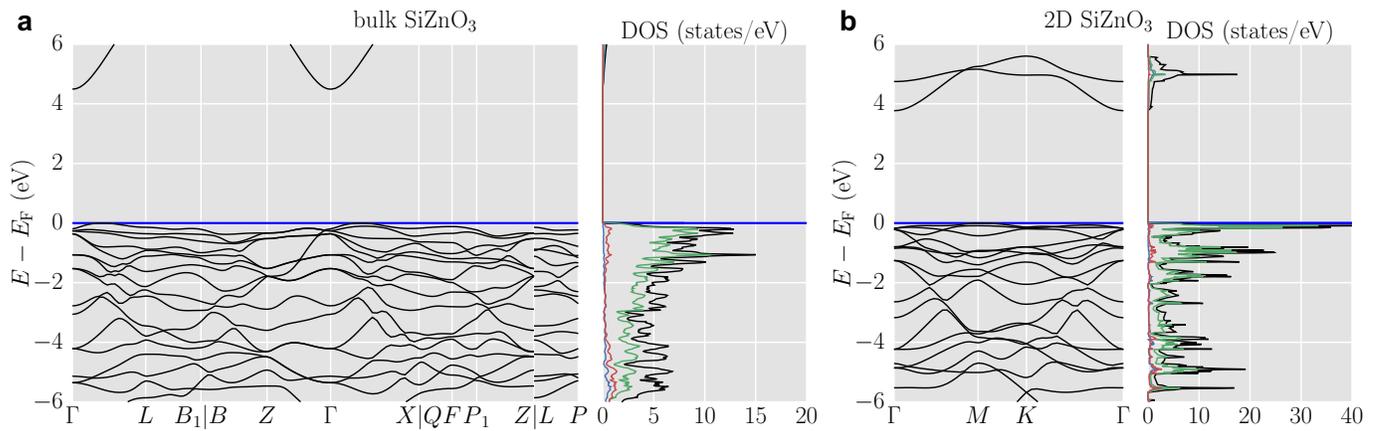

FIG. S30. Bandstructure and density of states for bulk (**a**) and 2D (**b**) SiZnO$_3$. The energies are aligned at the respective Fermi energy $E_F$.

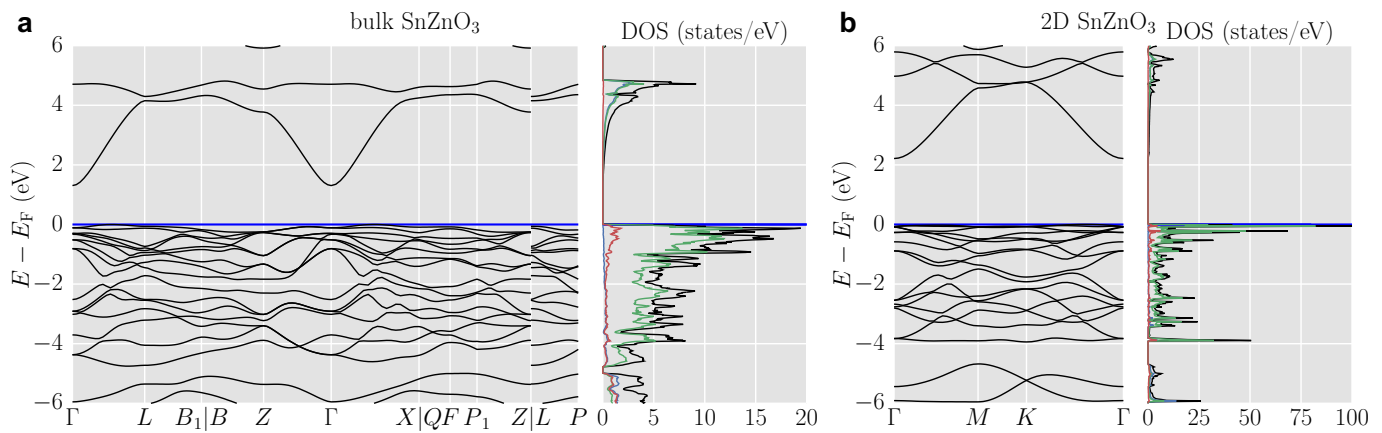

FIG. S31. Bandstructure and density of states for bulk (**a**) and 2D (**b**) SnZnO$_3$. The energies are aligned at the respective Fermi energy $E_F$.

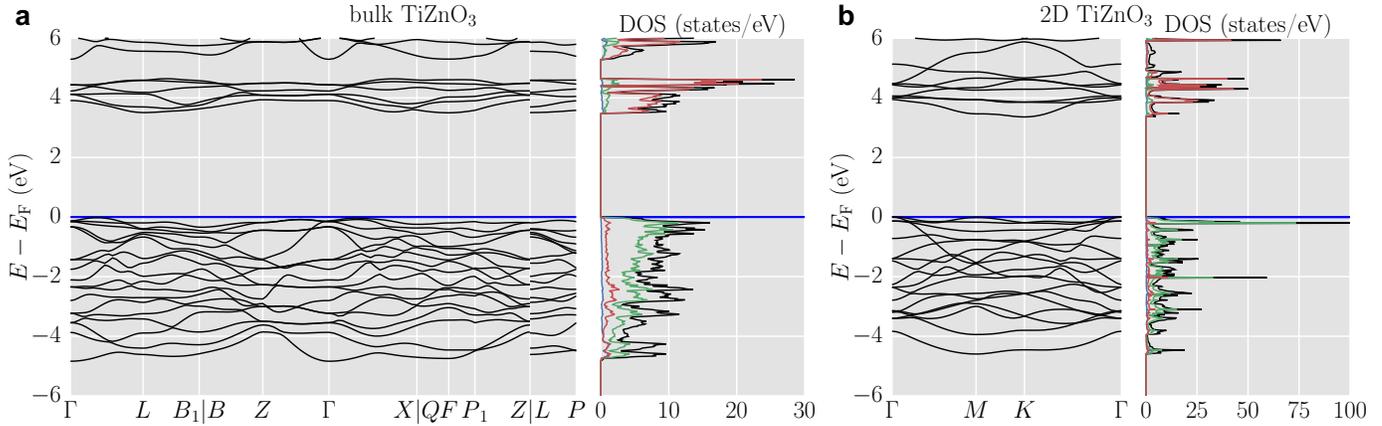

FIG. S32. Bandstructure and density of states for bulk (**a**) and 2D (**b**) TiZnO$_3$. The energies are aligned at the respective Fermi energy $E_F$.

## VI. Tables with numerical data

TABLE I: **Inter-facet binding energies for binaries.** Inter-facet binding energies for the binary systems calculated with different functionals from only a static electronic calculation, a relaxation of the ionic positions as well as also relaxing the (in-plane) cell parameters of the 2D materials. All values are in eV/Å$^2$.

| formula | LDA | | | PBE | | | PBE+$U$ | | | SCAN | | |
|---|---|---|---|---|---|---|---|---|---|---|---|---|
| | static | rel. ions | rel. ions and cell | static | rel. ions | rel. ions and cell | static | rel. ions | rel. ions and cell | static | rel. ions | rel. ions and cell |
| Al$_2$O$_3$ | 0.500 | 0.272 | 0.264 | 0.434 | 0.208 | 0.200 | 0.434 | 0.208 | 0.200 | 0.504 | 0.260 | 0.250 |
| Cr$_2$O$_3$ | 0.408 | 0.236 | 0.208 | 0.304 | 0.168 | 0.158 | 0.326 | 0.190 | 0.188 | 0.390 | 0.246 | 0.242 |
| Fe$_2$O$_3$ | 0.314 | 0.212 | 0.184 | 0.230 | 0.150 | 0.108 | 0.260 | 0.144 | 0.140 | 0.270 | 0.170 | 0.164 |
| Ga$_2$O$_3$ | 0.342 | 0.180 | 0.176 | 0.274 | 0.124 | 0.120 | 0.284 | 0.128 | 0.124 | 0.340 | 0.170 | 0.164 |
| In$_2$O$_3$ | 0.276 | 0.182 | 0.178 | 0.224 | 0.134 | 0.130 | 0.232 | 0.138 | 0.134 | 0.272 | 0.176 | 0.170 |
| Rh$_2$O$_3$ | 0.318 | 0.212 | 0.212 | 0.234 | 0.134 | 0.134 | 0.276 | 0.168 | 0.168 | 0.308 | 0.196 | 0.196 |
| Ti$_2$O$_3$ | 0.352 | 0.198 | 0.194 | 0.266 | 0.130 | 0.124 | 0.254 | 0.124 | 0.122 | 0.282 | 0.162 | 0.154 |
| V$_2$O$_3$ | 0.414 | 0.212 | 0.204 | 0.290 | 0.128 | 0.116 | 0.282 | 0.176 | 0.146 | 0.296 | 0.186 | 0.162 |

TABLE II: **Inter-facet binding energies for ternaries.** Inter-facet binding energies for the ternary systems calculated with PBE+$U$ and SCAN for fully relaxing all systems, *i.e.* the ionic positions and the (in-plane) cell parameters. The data for both the energetically preferred (more stab.) and the less preferred (less stab.) cation termination are given. All values are in eV/Å$^2$.

| formula | PBE+$U$ | | SCAN | | formula | PBE+$U$ | | SCAN | |
|---|---|---|---|---|---|---|---|---|---|
| | more stab. | less stab. | more stab. | less stab. | | more stab. | less stab. | more stab. | less stab. |
| AgBiO$_3$ | 0.018 | 0.116 | 0.034 | 0.078 | GeMnO$_3$ | 0.082 | 0.146 | 0.056 | 0.100 |
| BiNaO$_3$ | 0.028 | 0.210 | 0.040 | 0.234 | GeZnO$_3$ | 0.052 | 0.174 | 0.080 | 0.218 |
| CaSnO$_3$ | 0.090 | 0.190 | 0.116 | 0.248 | MgSiO$_3$ | 0.126 | 0.172 | 0.160 | 0.222 |
| CdGeO$_3$ | 0.066 | 0.142 | 0.096 | 0.190 | MgTiO$_3$ | 0.112 | 0.246 | 0.154 | 0.266 |
| CdTiO$_3$ | 0.072 | 0.230 | 0.108 | 0.248 | MnNiO$_3$ | 0.088 | 0.200 | 0.130 | 0.262 |
| CoMnO$_3$ | 0.084 | 0.194 | 0.130 | 0.200 | MnTiO$_3$ | 0.084 | 0.238 | 0.112 | 0.252 |
| CoTiO$_3$ | 0.082 | 0.246 | 0.112 | 0.270 | NiTiO$_3$ | 0.094 | 0.244 | 0.138 | 0.264 |
| CuVO$_3$ | 0.032 | 0.182 | 0.066 | 0.212 | SiZnO$_3$ | 0.062 | 0.198 | 0.092 | 0.240 |
| FeTiO$_3$ | 0.056 | 0.206 | 0.088 | 0.228 | SnZnO$_3$ | 0.040 | 0.204 | 0.066 | 0.268 |
| GeMgO$_3$ | 0.114 | 0.164 | 0.146 | 0.214 | TiZnO$_3$ | 0.054 | 0.254 | 0.094 | 0.272 |

TABLE III: **Structural properties for binaries.** In plane lattice constants and slab thicknesses for the relaxed binary 2D systems and as cut from the corresponding bulk structure for PBE+$U$ and SCAN. All values are in Å.

| formula | PBE+$U$ | | | | SCAN | | | |
|---|---|---|---|---|---|---|---|---|
| | $a_{2D}$ | $a_{bulk}$ | $d_{2D}$ | $d_{bulk}$ | $a_{2D}$ | $a_{bulk}$ | $d_{2D}$ | $d_{bulk}$ |
| Al$_2$O$_3$ | 4.922 | 4.807 | 2.721 | 3.877 | 4.864 | 4.749 | 2.725 | 3.839 |
| Cr$_2$O$_3$ | 5.145 | 5.060 | 3.112 | 4.206 | 5.023 | 4.943 | 3.146 | 4.188 |

TABLE III. (*continued*)

| formula | PBE+$U$ | | | | SCAN | | | |
|---|---|---|---|---|---|---|---|---|
| | $a_{2D}$ | $a_{bulk}$ | $d_{2D}$ | $d_{bulk}$ | $a_{2D}$ | $a_{bulk}$ | $d_{2D}$ | $d_{bulk}$ |
| Fe$_2$O$_3$ | 5.195 | 5.091 | 3.025 | 4.063 | 5.113 | 5.003 | 2.570 | 3.951 |
| Ga$_2$O$_3$ | 5.129 | 5.032 | 2.762 | 3.918 | 5.080 | 4.982 | 2.753 | 3.894 |
| In$_2$O$_3$ | 5.657 | 5.543 | 3.071 | 4.210 | 5.623 | 5.510 | 3.056 | 4.141 |
| Rh$_2$O$_3$ | 5.153 | 5.175 | 3.435 | 4.255 | 5.106 | 5.138 | 3.424 | 4.179 |
| Ti$_2$O$_3$ | 5.358 | 5.346 | 2.897 | 4.132 | 5.209 | 5.105 | 2.655 | 4.277 |
| V$_2$O$_3$ | 5.250 | 5.108 | 2.721 | 4.217 | 5.197 | 4.964 | 2.474 | 4.290 |

TABLE IV. **Structural properties for ternaries.** In plane lattice constants and slab thicknesses for the relaxed ternary 2D systems and as cut from the corresponding bulk structure for PBE+$U$ and SCAN for the energetically preferred cation termination. All values are in Å.

| formula | PBE+$U$ | | | | SCAN | | | |
|---|---|---|---|---|---|---|---|---|
| | $a_{2D}$ | $a_{bulk}$ | $d_{2D}$ | $d_{bulk}$ | $a_{2D}$ | $a_{bulk}$ | $d_{2D}$ | $d_{bulk}$ |
| AgBiO$_3$ | 5.802 | 5.729 | 3.254 | 4.312 | 5.774 | 5.700 | 2.977 | 4.092 |
| BiNaO$_3$ | 5.725 | 5.674 | 3.805 | 4.469 | 5.643 | 5.591 | 3.803 | 4.377 |
| CaSnO$_3$ | 5.659 | 5.637 | 3.678 | 4.277 | 5.516 | 5.509 | 3.696 | 4.216 |
| CdGeO$_3$ | 5.229 | 5.190 | 3.297 | 4.037 | 5.159 | 5.120 | 3.275 | 4.010 |
| CdTiO$_3$ | 5.394 | 5.337 | 3.279 | 4.103 | 5.300 | 5.250 | 3.261 | 4.074 |
| CoMnO$_3$ | 5.072 | 5.010 | 2.971 | 3.854 | 4.945 | 4.882 | 3.004 | 3.866 |
| CoTiO$_3$ | 5.247 | 5.153 | 2.960 | 4.007 | 5.164 | 5.044 | 2.732 | 3.961 |
| CuVO$_3$ | 5.101 | 4.997 | 2.966 | 4.014 | 5.030 | 4.918 | 2.877 | 4.001 |
| FeTiO$_3$ | 5.276 | 5.212 | 2.930 | 4.018 | 5.159 | 5.076 | 2.907 | 4.098 |
| GeMgO$_3$ | 5.084 | 5.013 | 3.039 | 3.922 | 5.014 | 4.943 | 3.043 | 3.880 |
| GeMnO$_3$ | 5.146 | 5.096 | 3.156 | 3.986 | 4.813 | 5.012 | 4.036 | 4.473 |
| GeZnO$_3$ | 5.047 | 4.989 | 2.774 | 3.707 | 5.026 | 4.973 | 2.801 | 3.660 |
| MgSiO$_3$ | 4.838 | 4.783 | 3.056 | 3.825 | 4.783 | 4.726 | 3.052 | 3.805 |
| MgTiO$_3$ | 5.242 | 5.140 | 3.031 | 4.040 | 5.150 | 5.050 | 3.031 | 4.007 |
| MnNiO$_3$ | 5.038 | 4.977 | 3.032 | 3.943 | 4.919 | 4.867 | 3.080 | 3.937 |
| MnTiO$_3$ | 5.304 | 5.227 | 3.105 | 4.070 | 5.202 | 5.114 | 2.986 | 4.040 |
| NiTiO$_3$ | 5.215 | 5.118 | 3.050 | 4.139 | 5.111 | 5.018 | 3.060 | 4.124 |
| SiZnO$_3$ | 4.803 | 4.757 | 2.833 | 3.626 | 4.792 | 4.744 | 2.900 | 3.668 |
| SnZnO$_3$ | 5.475 | 5.364 | 2.738 | 3.933 | 5.391 | 5.301 | 2.740 | 3.812 |
| TiZnO$_3$ | 5.205 | 5.112 | 2.790 | 3.859 | 5.164 | 5.072 | 2.839 | 3.864 |

TABLE V. **Magnetic properties for binaries with SCAN.** Energetically preferred magnetic ordering type, energy difference $\Delta E = E_{AFM} - E_{FM}$ between AFM and FM ordering in eV/formula unit, and absolute magnetic moments for the different sublattices in $\mu_B$ for the bulk and corresponding [001] facets as calculated with SCAN. NM stands for non-magnetic.

| formula | mag. order type | | $\Delta E$ | | $\mu_{bulk}$ | $\mu_{2D}$ | |
|---|---|---|---|---|---|---|---|
| | bulk | 2D | bulk | 2D | | inner | outer |
| Cr$_2$O$_3$ | AFM | AFM | −0.150 | −0.044 | 2.73 | 2.72 | 2.73 |
| Fe$_2$O$_3$ | AFM | AFM | −0.801 | −0.323 | 3.93 | 3.68 | 3.57 |
| Ti$_2$O$_3$ | NM | FM | - | 0.118 | - | 0.75 | 0.15 |
| V$_2$O$_3$ | FM | AFM | 0.012 | −0.247 | 1.89 | 2.21 | 0.03 |

TABLE VI. **Magnetic properties for ternaries with SCAN.** Energetically preferred magnetic ordering type, energy difference $\Delta E = E_{AFM} - E_{FM}$ between AFM and FM ordering in eV/formula unit, and absolute magnetic moments for the inner and outer magnetic ions of the slab in $\mu_B$ for the bulk and corresponding [001] facets as calculated with SCAN. The $A$ element is the first cation species in the formula while the $B$ element corresponds to the second. The terminating elements of the slabs (outer ions) are underlined.

| formula | mag. order type | | $\Delta E$ | | $\mu_{bulk}$ | | $\mu_{2D}$ | |
|---|---|---|---|---|---|---|---|---|
| | bulk | 2D | bulk | 2D | $A$ el. | $B$ el. | inner | outer |
| <u>Co</u>MnO$_3$ | AFM | AFM | −0.386 | −0.338 | 2.50 | 2.62 | 2.92 | 2.48 |
| <u>Co</u>TiO$_3$ | FM | AFM | 0.014 | −0.172 | 2.58 | 0.09 | 0.00 | 2.43 |
| <u>Fe</u>TiO$_3$ | FM | AFM | 0.102 | −0.018 | 3.72 | 0.08 | 0.02 | 3.53 |
| <u>Ge</u>MnO$_3$ | AFM | FM | −0.061 | 0.004 | 0.01 | 4.49 | 2.71 | 0.02 |
| <u>Mn</u>NiO$_3$ | AFM | AFM | −0.244 | −0.279 | 2.63 | 1.56 | 2.73 | 1.53 |
| <u>Mn</u>TiO$_3$ | AFM | AFM | −0.049 | −0.038 | 4.42 | 0.02 | 0.00 | 4.36 |
| <u>Ni</u>TiO$_3$ | AFM | AFM | −0.002 | −0.003 | 1.64 | 0.03 | 0.01 | 1.61 |

[1] R. Zacharia, H. Ulbricht, and T. Hertel, *Interlayer cohesive energy of graphite from thermal desorption of polyaromatic hydrocarbons*, Phys. Rev. B **69**, 155406 (2004).

[2] N. Mounet, M. Gibertini, P. Schwaller, D. Campi, A. Merkys, A. Marrazzo, T. Sohier, I. E. Castelli, A. Cepellotti, G. Pizzi, and N. Marzari, *Two-dimensional materials from high-throughput computational exfoliation of experimentally known compounds*, Nat. Nanotechnol. **13**, 246–252 (2018).

[3] Y. Zhang, J. W. Furness, B. Xiao, and J. Sun, *Subtlety of TiO2 phase stability: Reliability of the density functional theory predictions and persistence of the self-interaction error*, J. Chem. Phys. **150**, 014105 (2019).